%% file: main.tex
\documentclass[twocolumn]{article}
\usepackage{parskip}
\usepackage{amssymb}
\usepackage{amsmath}
\usepackage{fullpage}
\usepackage{authblk}
\usepackage{blindtext}
\usepackage{siunitx}
\usepackage{float}
\usepackage{caption}
\usepackage[superscript,nomove]{cite}
\usepackage{hyperref}
\usepackage[normalem]{ulem}
\usepackage{graphicx}
\usepackage{cancel}
\usepackage{pifont}
\usepackage{diagbox}
\usepackage[percent]{overpic}
\usepackage{tikz}
\usetikzlibrary{arrows.meta,positioning,calc,fit}
\usetikzlibrary{shapes.geometric}
\usepackage{widetext}
\usepackage{lipsum}
\usepackage{booktabs}
\usepackage{tabularx}
\usepackage{subcaption}
\usepackage{caption}
\usepackage{subcaption}
\usepackage{stfloats}
\usepackage{orcidlink}
\newcommand{\iu}{\mathrm{i}\mkern1mu}

\newcommand{\CompleteGraph}[2]{%
  \begin{tikzpicture}[every node/.style={font=\small}]
    \def\n{#1}
    \pgfmathsetmacro{\R}{#2}

    \foreach \i in {1,...,\n} {
      \pgfmathsetmacro{\ang}{90-360*(\i-1)/\n}
      \coordinate (v\i) at ({\R*cos(\ang)},{\R*sin(\ang)});
      \node[circle, fill, inner sep=1.6pt,
            label={[font=\small] \ang:${\i}$}] at (v\i) {};
    }

    \foreach \i in {1,...,\n} {
      \pgfmathtruncatemacro{\jstart}{\i+1}
      \ifnum\i<\n
        \foreach \j in {\jstart,...,\n} {
          \draw (v\i) -- (v\j);
        }
      \fi
    }
  \end{tikzpicture}%
}

\newcommand{\CompleteGraphMath}[2]{%
  \vcenter{\hbox{\CompleteGraph{#1}{#2}}}
}

\makeatletter \renewcommand{\@citess}[1]{\textsuperscript{[#1]}} \makeatother

\title{Tensor cluster expansion as an efficient surrogate for multi-ACE}
\author[1, 2]{\small Jacob Jeffries \orcidlink{0000-0002-8852-6765} \thanks{jwjeffr@g.clemson.edu}}
\author[1, 3]{Enrique Martinez \orcidlink{0000-0002-2690-2622}}
\author[2]{Eva Zarkadoula \orcidlink{0000-0002-6886-9664
}\thanks{zarkadoulae@ornl.gov}}
\affil[1]{Department of Materials Science and Engineering, Clemson University, Clemson, SC 29634, USA}
\affil[2]{Center for Nanophase Materials Sciences, Oak Ridge National Laboratory, Oak Ridge, TN 37831, USA}
\affil[3]{School of Automotive and Mechanical Engineering, Clemson University, Clemson, SC 29634, USA}
\date{\small \today}

\renewenvironment{abstract}
 {\quotation\small\noindent\rule{\linewidth}{.5pt}\par\smallskip
  {\centering\bfseries\abstractname\par}\medskip}
 {\par\noindent\rule{\linewidth}{.5pt}\endquotation}

\begin{document}

\twocolumn[
  \begin{@twocolumnfalse}
  \maketitle
    \begin{abstract}
        In this work, we generalize an on-lattice tensor cluster expansion (TCE) framework to an arbitrary correlation order. We show that these generalized tensor cluster expansion features span a subspace of MACE features, suggesting that TCE features can sufficiently pre-screen MACE features for chemical diversity in alloyed transition metal dichalcogenide (TMD) monolayers. We show that, using maximum-entropy subset sampling, we can empirically recover chemical diversity from MACE features using a cheaper TCE representation. We then fine-tune a multi-ACE (MACE) potential with both subsets, and use this potential to predict temperature-dependent vibrational properties in alloyed TMD monolayers with varying chemical ordering. In our approach, TCE features can be used as cheap surrogate features for pre-screening chemically rich configurations using MACE features, allowing for the more efficient finetuning of atomic cluster expansion (ACE) potentials, thus achieving a tractable prediction of finite-temperature alloy thermodynamics on modest hardware.
    \end{abstract}
  \vspace{0.5cm}
  \end{@twocolumnfalse}
]

\begin{figure*}[!b]
\footnotesize
\noindent
\rule{\linewidth}{.5pt}
Notice: This manuscript has been coauthored by UT-Battelle, LLC, under contract DE-AC05-00OR22725 with the US Department of Energy (DOE). The US government retains and the publisher, by accepting the article for publication, acknowledges that the US government retains a nonexclusive, paid-up, irrevocable, worldwide license to publish or reproduce the published form of this manuscript, or allow others to do so, for US government purposes. DOE will provide public access to these results of federally sponsored research in accordance with the DOE Public Access Plan (\url{https://www.energy.gov/doe-public-access-plan}).
\end{figure*}

\input{sections/introduction}
\input{sections/theory-and-methods}
\input{sections/results}
\input{sections/discussion}
\input{sections/limitations}
\input{sections/conclusions}
\input{sections/data_availability}
\input{sections/acknowledgements}
\input{sections/disclaimer}

\bibliography{bibfile.bib}

\input{sections/appendix}

\end{document}

%% file: sections/introduction.tex
\section{Introduction}

Machine learning interatomic potentials (MLIPs) have shown recent successes in large-scale atomistic simulations of various materials systems \cite{deringer2019machine, mueller2020machine, mishin2021machine}. In particular, the atomic cluster expansion (ACE) formalism \cite{drautz2019atomic} extends the Ising/spin-like cluster expansion techniques to construct a complete basis \cite{dusson2022atomic} for describing local chemical environments, while multi-ACE (MACE) \cite{batatia2022mace, batatia2025design} extends the ACE potential to efficiently parametrize many-body clusters. Furthermore, recent work has demonstrated that MACE potentials can reproduce vibrational properties and Raman spectra across transition-metal dichalcogenide (TMD) alloys, extending first-principles calculations to larger cells and broader configurational sampling \cite{siddiqui2024machine}.

In general, training an MLIP from the ground up is computationally expensive. An emerging alternative is to fine-tune foundation models, leveraging the extensive datasets used during pretraining while adapting the models to make quantitatively accurate predictions for specialized applications, including heterogeneous catalysis, point-defect migration, dislocation mechanics, solid-state phase transitions, and battery electrolyte dynamics. For example, Batatia et al. demonstrated that fine-tuning the MACE-MP-0 foundation model using relatively small application-specific datasets could substantially improve predictions of catalytic reaction barriers, defect formation and migration energies, and electrolyte transport properties\cite{batatia2025foundation}.

Primarily, the computational bottleneck is within the generation of first-principles data for such application. Although fine-tuning has significantly reduced the required amount of new first-principles data \cite{kaur2025data}, the generation of each single data point still requires specialized hardware, and is very computationally expensive. This is especially problematic for the study of complex systems, e.g. concentrated alloys, where the prohibitively large design space necessitates the need for a relatively large number of data points. 

Strategic selection of a training set is therefore a necessity for efficiently screening new alloy systems. The common strategy is to generate a superset of configurations $\mathcal{X}$, then select a subset $\mathcal{X}^*\subseteq\mathcal{X}$ that maximizes some global footprint for chemical diversity. These footprints are usually computed using feature vectors that describe atomic configurations, which, importantly, can be computed independent of \textit{ab-initio} data, i.e. solely from atomic positions and species \cite{karabin2020entropy, qi2024robust, schwalbe2025model, poul2025automated}. Particularly relevant to this work, Arora et al. used bond counts to stratify configurations of ternary, quaternary, and quinary BCC alloys, using this stratification to pre-screen for chemical complexity prior to performing density functional theory (DFT) calculations \cite{arora2026bond}. However, as noted by the authors, many-body terms were not included within the descriptor space, meaning that their approach is not suitable for screening for many-body effects. Furthermore, this descriptor is agnostic of lattice relaxation effects, which is potentially problematic for pre-screening configurations with large local lattice distortion.

In the ideal case, the feature vector used to describe the atomic configuration is the same feature vector that is used as input to the MLIP. This ensures that the diversity maximized in the training set is mappable to the diversity of the outputs of the MLIP. Using the MLIP feature vector, in contrast to Arora et al. \cite{arora2026bond}, would furthermore induce expressivity in regard to lattice relaxation effects. In the case where the set used to finetune the MLIP is of modest size, the calculation of the feature vector within the diversity sampling routine described above is tractable. However, for expensive feature vectors, this can become untractable for large training sets. This problem is particularly unavoidable for multicomponent systems, where the number of chemically complex environments increases combinatorially with the number of components in the system of interest.

Therefore, sufficient sampling diversity is of utmost importance for the screening of novel, chemically complex materials, including bulk high-entropy alloys for clean energy applications \cite{xia2016phase, granberg2016mechanism, lu2016enhancing} and, more recently, concentrated two-dimensional TMDs such as the alloyed monolayer $(\text{V}\text{Nb}\text{Mo}\text{Ta}\text{W})\text{S}_2$ for $\text{C}\text{O}_2$ electroreduction \cite{cavin20212d} and, more generally, tuneable vacancy stability \cite{gao2025point}. Beyond electrocatalysis, alloyed TMDs have demonstrated potential for optoelectronic applications, including broadband photodetection and photovoltaic devices, through composition-dependent optical properties and defect engineering \cite{hou2020hydrothermally, sharma2020spectroscopic}. Chalcogen substitution in particular has enabled the tuning of valley polarization and photocurrent generation in $\mathrm{Mo}(\mathrm{S},\mathrm{Se})_2$ monolayers, while simultaneous metal and chalcogen substitution has been explored for enhancing valley polarization at room temperature \cite{nayak2024valley, maity2024tuning}. More broadly, alloying provides a route to engineering spin-orbit interactions and accessing competing structural and topological phases, further motivating the exploration of chemically complex TMD monolayers \cite{wang2015spin, oliver2017structural}.

However, high-entropy alloying as a strategy presents a major challenge, in that compositional complexities combinatorially explode the number of configurations needed to sufficiently sample the chemical diversity of $\mathcal{X}$. Consequently, it is advantageous to select a computationally cost-effective, surrogate pre-screener to compute feature vectors, rather than compute the whole MLIP feature vector for a set of atomic configurations.

In this work, we show that the tensor cluster expansion (TCE) framework yields feature vectors that span a subspace of MACE feature vectors. We then show that, using greedy maximum-entropy subset selection, TCE feature vectors act as an efficient pre-screening surrogate for configurational diversity as measured by MACE feature vectors in model concentrated TMD monolayers. Lastly, we show that fine-tuning using this computationally efficient pre-screening method quantitatively predicts vibrational spectra of model TMDs.

%% file: sections/theory-and-methods.tex
\section{Theory}

To characterize chemical complexity in the material, we seek a feature vector that serves as a chemical fingerprint. To capture this complexity, we  compute $m$-point correlation functions corresponding to short-range chemical ordering, i.e. a cluster expansion method. We, thus, generalize the tensor cluster expansion (TCE) method for $m$-body terms. TCE computes two- and three-body cluster counts via tensor products between topological tensors and occupation tensors, respectively, defining the nearest-neighbor topology and occupancy within a static lattice:

\begin{equation}\label{eq:tce-old}
    \begin{aligned}
        N_{\alpha\beta}^{(n)} &= A_{ij}^{(n)}X_{i\alpha}X_{j\beta} \\
        M_{\alpha\beta\gamma}^{(n)} &= B_{ijk}^{(n)}X_{i\alpha}X_{j\beta}X_{k\gamma}
    \end{aligned}
\end{equation}

where Latin indices denote points within the lattice, Greek indices denote chemical species within the lattice, $n$ denotes topological orders for different clusters, $\mathbf{X}$ denotes occupational tensors within the lattice so that:

\begin{equation}
    X_{i\alpha} = \begin{cases}
        1 & \text{site $i$ is occupied by type $\alpha$} \\
        0 & \text{else}
    \end{cases}
\end{equation}

Last, $\mathbf{A}$ and $\mathbf{B}$ denote topological tensors corresponding to two- and three-body clusters:

\begin{equation}\label{eq:topology-tensors-old}
    \begin{aligned}
        A_{ij}^{(n)} &= \begin{cases}
            1 & \text{sites $(i, j)$ are $n$'th nearest neighbors} \\
            0 & \text{else}
        \end{cases} \\
        B_{ijk}^{(n)} &= \begin{cases}
            1 & \text{sites $(i, j, k)$ are $n$'th order triplets} \\
            0 & \text{else}
        \end{cases}
    \end{aligned}
\end{equation}

so that $\mathbf{N}$ and $\mathbf{M}$ defined in Eq.~\eqref{eq:tce-old} store the counts of each cluster type, e.g. the number of first nearest neighbor $\alpha$-$\beta$ interactions, or the number of $\alpha$-$\beta$-$\gamma$ equilateral triangles in a highly symmetric structure like a face-centered cubic (FCC) solid. Here, the tensor $\mathbf{A}$ can be built by a neighbor list, and $\mathbf{B}$ can be built as a connected tensor product of copies of $\mathbf{A}$. For example, in the case of FCC, this tensor product is:

\begin{equation}\label{eq:tensor-product-legs}
    B_{ijk}^{(1)} = A_{ij}^{(1)}A_{jk}^{(1)}A_{ki}^{(1)}
\end{equation}

i.e. each pair of lower indices corresponds to a pair of neighbors, and each upper label corresponds to the type of interaction order between those neighbors \cite{jeffries2026generalized}.

In this work, we generalize these expressions to $m$-bodies, expanding our previous work where we used only the $2$ and $3$ body terms ~\cite{jeffries2026generalized}. For $m = 3$, the connected tensor product to compute $\mathbf{B}$ in Eq.~\eqref{eq:tensor-product-legs} is induced by the fully connected graph of size $3$, commonly denoted as $K_3$:

\begin{equation}
    K_3 = \CompleteGraphMath{3}{0.5}
\end{equation}

so that each pair of lower indices in connected tensor products like Eq.~\eqref{eq:tensor-product-legs} corresponds to an edge of $K_3$:

\begin{equation}
    \begin{aligned}
        B_{i_1i_2i_3}^{(\ell_{12}, \ell_{23}, \ell_{13})} &= A_{i_1i_2}^{(\ell_{12})} A_{i_2i_3}^{(\ell_{23})} A_{i_1i_3}^{(\ell_{13})}\\ 
        &= \prod_{\{r, s\}\in E(K_3)} A_{i_ri_s}^{(\ell_{rs})} \\
        B_{i_1i_2i_3}^{[\ell]} &= \bigvee_{\ell'\in[\ell]} B_{i_1i_2i_3}^{(\ell')}
    \end{aligned}
\end{equation}

where $[\ell] = S_3\cdot\ell$ denotes the orbit of $\ell$ under the action of $S_3$ (the permutation group on $3$ elements) and $E(G)$ denotes the edge set of a graph $G$. We then symmetrize so that the final tensor $\mathbf{B}^{[\ell]}$ is fully symmetric and remains Boolean, i.e. $B_{i_1i_2i_3}^{[\ell]}\in\{0, 1\}$.

We then generalize the topological tensors to include edges from the graph $K_m$ to represent $m$-body clusters, rather than just up to $K_3$:

\begin{equation}\label{eq:top-from-graph}
    A^{[\ell]}_{i_1\cdots i_m} = \bigvee_{\ell'\in[\ell]}\prod_{\{r, s\}\in E(K_m)} A_{i_r i_s}^{(\ell_{rs}')}
\end{equation}

where $[\ell] = S_m\cdot\ell$, $K_m = (V_m, E_m)$ denotes a labeled, fully connected, undirected graph with vertices $V_m$ and edges $E_m$:

\begin{equation}
    \begin{aligned}
        V_m &= \{1, \cdots, m \} \\
        E_m &= \left\{ \{u, v\}\;|\;(u, v)\in V_m\times V_m \text{ and } u \neq v\right\}
    \end{aligned}
\end{equation}

For example, for $m = 4$ and $m = 5$:

\begin{equation}
    \begin{aligned}
        K_4 = \CompleteGraphMath{4}{0.5} \qquad K_5 = \CompleteGraphMath{5}{0.5}
    \end{aligned}
\end{equation}

Then, using index notation, the tensor $\mathbf{A}$ can be interpreted as a generalized topological tensor:

\begin{equation}
    A_{\boldsymbol{i}}^{[\ell]} = \begin{cases}
        1 & \text{sites $\boldsymbol{i}$ form a cluster with topology $[\ell]$} \\
        0 & \text{else}
    \end{cases}
\end{equation}

The tensor $\mathbf{A}^{[\ell]}$ is notably symmetric, i.e. $A_{i_1\cdots i_m}^{[\ell]} = A_{(i_1\cdots i_m)}^{[\ell]}$, which is provable from Eq.~\eqref{eq:top-from-graph} by noting that both $K_m$ and $[\ell]$ are invariant under the action $S_m$. Furthermore, the number of clusters with occupying types $\boldsymbol{\alpha} = (\alpha_1, \cdots, \alpha_m)$ and geometry $[\ell]$ is:

\begin{equation}\label{eq:tensor-contraction}
    N_{\alpha_1\cdots\alpha_m}^{[\ell]} = A_{i_1\cdots i_m}^{[\ell]} \prod_{k=1}^m X_{i_k\alpha_k}
\end{equation}

which is notably also symmetric, i.e. $N_{\alpha_1\cdots\alpha_m}^{[\ell]} = N_{(\alpha_1\cdots\alpha_m)}^{[\ell]}$, yielding an effective Hamiltonian:

\begin{equation}
    \begin{aligned}
        \mathcal{H}_\text{eff} &= \sum_{[\ell]\in\Lambda} \frac{1}{m!}\varepsilon_{\alpha_1\cdots\alpha_m}^{[\ell]}N_{\alpha_1\cdots\alpha_m}^{[\ell]}
    \end{aligned}
\end{equation}

and a generalized feature vector:

\begin{equation}\label{eq:tce-feat-vec}
    \mathbf{t} = \bigoplus_{[\ell]\in\Lambda} \operatorname{vec}\left(N_{\alpha_1\cdots\alpha_m}^{[\ell]}\right)
\end{equation}

where $\Lambda$ is an \textit{a-priori} chosen set of topologies, $\oplus$ denotes a direct sum over cluster subspaces (equivalently concatenation), and $m$ is the body-order corresponding to the topological label $[\ell]$:

\begin{equation}
    |\ell| = \binom{m}{2} \implies m = \frac{1+\sqrt{1+8|\ell|}}{2}
\end{equation}

The only inputs to computing $\mathbf{t}$ are then the lattice positions and occupying types to respectively define $\mathbf{A}$ and $\mathbf{X}$, as well as the chosen topologies $\Lambda$. Each cluster corresponds solely to a topological label $\ell$, labeling each edge within the cluster by its neighbor order. For example, a second-nearest-neighbor interaction can be labeled with $\ell = 2$, and a nearest-neighbor triangle is labeled with $\ell = (1, 1, 1)$. Then, the tensor contraction in Eq.~\eqref{eq:tensor-contraction} computes the number of species-decorated clusters with the topological label $\ell$, as well as all other topological labels that are symmetrically equivalent to it, namely the set $[\ell] = S_m\cdot\ell$. Furthermore, $\mathbf{A}$ can be computed once per set of lattice positions, i.e. $\mathbf{A}$ does not need to be recomputed if atom types are permuted within one lattice.

Furthermore, similar to our work in Ref.~\cite{jeffries2026generalized}, this Hamiltonian has the advantage that we can efficiently compute the change in cluster counts from some two-particle swap in Eq.~\eqref{eq:tensor-contraction} by simply truncating the contraction to only include sites within those swaps, i.e.:

\begin{equation}\label{eq:swap-shortcut}
    \begin{aligned}
        \Delta \tilde{N}_{\alpha_1\cdots\alpha_m}^{[\ell]} &= \sum_{i_1\in\operatorname{supp}\pi} A_{i_1\cdots i_m}^{[\ell]}X_{i_1\alpha_1}'\prod_{k=2}^m X_{i_k\alpha_k}'\\
        &-\sum_{i_1\in\operatorname{supp}\pi} A_{i_1\cdots i_m}^{[\ell]}X_{i_1\alpha_1}\prod_{k=2}^m X_{i_k\alpha_k}\\
        \Delta N_{\alpha_1\cdots\alpha_m}^{[\ell]} &= m\Delta \tilde{N}_{(\alpha_1\cdots\alpha_m)}^{[\ell]}        
    \end{aligned}
\end{equation}

where $\pi\in S_N$ is a $2$-cycle acting on $N$ lattice sites, $S_N$ is the permutation group acting on the $N$ lattice sites in the system, $\operatorname{supp}\pi$ is the support of that permutation, and $\mathbf{X}'=\pi\mathbf{X}$ is the permuted state tensor, and parenthesis around indices denote symmetrization with respect to those indices. Then, since any permutation can be decomposed into $2$-cycles, Eq.~\eqref{eq:swap-shortcut} can be then used iteratively for more complex permutations, and more generally any Monte Carlo move that preserves global composition. Then, since the Hamiltonian is linear in $\mathbf{N}$, we can efficiently compute the change in energy from the swap:

\begin{equation}
    \Delta E(\mathbf{X}\to\pi\mathbf{X}) = \sum_{[\ell]\in\Lambda}\varepsilon_{\alpha_1\cdots\alpha_m}^{[\ell]}\Delta N_{\alpha_1\cdots\alpha_m}^{[\ell]}    
\end{equation}

i.e., the model explicitly encodes local atomic energies, and linearity of the model allows us to efficiently evaluate energy differences by ignoring bonds far away from where the swap(s) occur, allowing for very efficient canonical Monte Carlo simulations in which any particle swap can be represented by a permutation on lattice sites $\pi\in S_N$, as well as difference training in which we learn the interaction coefficients from difference in cluster counts, rather than raw cluster counts, similar to Ref.~\cite{musa2025accelerating}. A similar formula can be applied for single-site transmutations by changing the summation domain to include the transmutated lattice sites, also inducing efficient, topology-aware grand canonical Monte Carlo simulations up to arbitrary correlation order.

We furthermore find that the TCE features derived here are contained in a rotationally-invariant feature span of MACE features, of which we outline a proof for in Sec.~\ref{sec:proof-outline}, with a full proof available in Sec.~\ref{sec:full-proof}. The key observation is that, on a fixed lattice, the edges between neighbors can be recovered exactly from a finite set of root-centered angular configurations.

\section{Methods}\label{sec:methods}

\begin{figure*}[t]
    \centering
    \includegraphics[width=\linewidth]{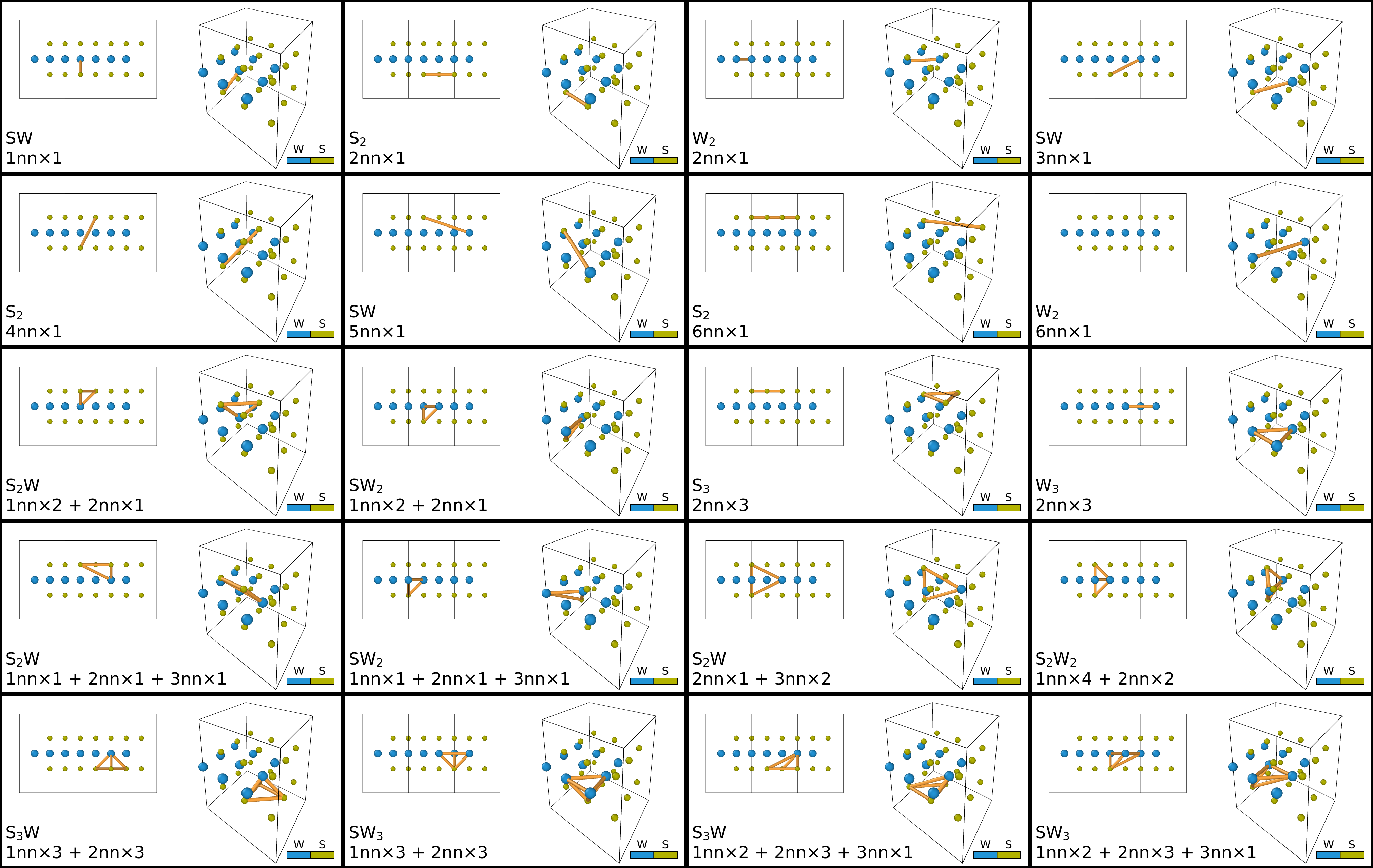}
    \caption{Topological feature set $\Lambda$ used for decorating $\text{W}\text{S}_2$-like structures. Each atom type represents a subset of possible atom types occupying that site, i.e. $\text{W}$ is a placeholder for a transition metal and $\text{S}$ is a placeholder for a chalcogenide, visualized with Open Visualization Tool (OVITO) \cite{stukowski2010visualization}.}
    \label{fig:ws2-clusters}
\end{figure*}

For the TCE feature vector, we compute a normalized version of Eq.~\eqref{eq:tce-feat-vec} for clusters in the high entropy TMD system $(\text{W},\text{Mo},\text{Ta})(\text{S},\text{Se},\text{Te})_2$ with clusters visualized in Fig.~\ref{fig:ws2-clusters}:

\begin{equation}
    \tilde{\mathbf{t}} = \bigoplus_{[\ell]\in\Lambda} \operatorname{vec}\left( N_{\alpha_1\cdots\alpha_m}^{[\ell]} / \sum_{i_1 \cdots i_m} A_{i_1\cdots i_m}^{[\ell]} \right)
\end{equation}

which effectively normalizes by both system size as well as the size of a given cluster subspace, in order to stabilize the covariance matrix. For example, for the two-body terms in a single-sublattice system:

\begin{equation}
    \sum_{ij}A_{ij}^{(1)} = Nz
\end{equation}

where $z$ is the first-nearest coordination number of the solid and $N$ is the number of atoms in the solid, so that $N_{\alpha\beta}^{(1)}/(Nz)$ denotes the proportion of first nearest-neighbor pairs that contain $\alpha$ and $\beta$ types. We can also compute the MACE feature $\tilde{\mathbf{t}} = \phi_\text{MACE}(x)$ by averaging per-atom descriptors, and compare.

For a given set of configurations indexed by $\mathcal{X}$, we can define the total entropy:

\begin{equation}
    S(\mathcal{X}) = \frac{1}{2}\log\det \left[ \sum_{x\in\mathcal{X}} \left(\tilde{\mathbf{t}}_x - \boldsymbol{\mu}\right)\otimes\left(\tilde{\mathbf{t}}_x - \boldsymbol{\mu}\right) \right]
\end{equation}

where $\boldsymbol{\mu}$ is the mean of the dataset:

\begin{equation}
    \boldsymbol{\mu} = \frac{1}{|\mathcal{X}|}\sum_{x\in\mathcal{X}}\tilde{\mathbf{t}}_x
\end{equation}

$\otimes$ denotes an outer product, i.e. $(\mathbf{x}\otimes \mathbf{y})_{ij} = x_i y_j$, $\sum_{x\in\mathcal{X}} \left(\tilde{\mathbf{t}}_x - \boldsymbol{\mu}\right)\otimes\left(\tilde{\mathbf{t}}_x - \boldsymbol{\mu}\right)$ is the covariance of the dataset. For brevity, we let $\mathbf{w}_x = \tilde{\mathbf{t}}_x - \boldsymbol{\mu}$. We then seek to find the optimal subset of size $k$ that maximizes this entropy:

\begin{equation}
    \mathcal{X}^* = \arg\max_{\substack{\mathcal{X}^*\subseteq\mathcal{X} \\ |\mathcal{X}^*| = k}}\log\det \left[ \sum_{x\in\mathcal{X}^*} \mathbf{w}_x\otimes\mathbf{w}_x \right]
\end{equation}

We can do this greedily by starting with some initial subset $\mathcal{X}_0$. Then, we can select the next configuration $y\in\mathcal{X}\setminus \mathcal{X}_0$ which maximizes the entropy gain \cite{krause2008near, sharma2015greedy}:

\begin{equation}
    \begin{aligned}
        S(\mathcal{X}_0\cup\{y\}) - S(\mathcal{X}_0) = \frac{1}{2}\log\frac{\det \boldsymbol{\Sigma}(\mathcal{X}_0\cup\{y\})}{\det\boldsymbol{\Sigma}(\mathcal{X}_0)}
    \end{aligned}
\end{equation}

where $\boldsymbol{\Sigma}(\mathcal{X}^*)$ is the covariance matrix:

\begin{equation}
    \begin{aligned}
        \boldsymbol{\Sigma}(\mathcal{X}_0) &= \sum_{x\in\mathcal{X}_0} \mathbf{w}_x\otimes\mathbf{w}_x \\
        \boldsymbol{\Sigma}(\mathcal{X}_0\cup\{y\}) &= \boldsymbol{\Sigma}(\mathcal{X}_0) + \mathbf{w}_y\otimes\mathbf{w}_y
    \end{aligned}
\end{equation}

Then, by the matrix determinant lemma, the entropy gain from adding sample $y$ to subset $\mathcal{X}_0$ is proportional to:

\begin{equation}
    \log\frac{\det \boldsymbol{\Sigma}(\mathcal{X}_0\cup\{y\})}{\det\boldsymbol{\Sigma}(\mathcal{X}_0)} = \log\left(1 + \mathbf{w}_y^\intercal \boldsymbol{\Sigma}(\mathcal{X}_0)^{-1}\mathbf{w}_y\right)
\end{equation}

We can then recursively define this procedure with a starting subset $\mathcal{X}_0$ and an initial covariance $\boldsymbol{\Sigma}_0$, namely:

\begin{equation}
    \begin{aligned}
        y_n &= \arg\max_{y\in\mathcal{X}\setminus\mathcal{X}_n} \log \left(1 + \mathbf{w}_y^\intercal\boldsymbol{\Sigma}_n^{-1}\mathbf{w}_y\right) \\
        \mathcal{X}_{n+1} &= \mathcal{X}_n\cup \{y_n\} \\
        \boldsymbol{\Sigma}_{n+1} &= \boldsymbol{\Sigma}_n + \mathbf{w}_{y_n}\otimes\mathbf{w}_{y_n}
    \end{aligned}
\end{equation}

For larger subsets, the inversion of $\boldsymbol{\Sigma}_n$ is very computationally expensive. However, one can avoid recomputing the full inverse via the Woodbury matrix identity \cite{higham2002accuracy}. We can iteratively perform this procedure for the TCE feature vector as well as the MACE feature vector, generating the respective subsets $\mathcal{X}_\text{TCE}$ and $\mathcal{X}_\text{MACE}$. We can measure the overlap between the two subsets as the size of their intersection:

\begin{equation}
    I = \left|\mathcal{X}_\text{TCE}\cap \mathcal{X}_\text{MACE}\right|
\end{equation}

where $|\mathcal{X}_\text{TCE}| = |\mathcal{X}_\text{MACE}| = k$. In the ideal case, where the two sets are equal, $I = k$. In the least ideal case, where the two subsets are statistically unrelated (i.e. all overlap is by random chance), $I = k^2/n$, where $n = \left|\mathcal{X}\right|$.

Using ASE \cite{hjorth2017atomic}, we generate 100 high entropy, random TMD monolayer solutions with $(\text{W}_{1-x-y}\text{Mo}_x\text{Ta}_y)(\text{S}_{1-z-w}\text{Se}_z\text{Te}_w)_2$. While generating each sample, we randomly select a subset of metals and chalcogens, and then randomly select compositions within those subsets, to ensure that our dataset not only contains the full senary system, but also the relevant chemical subsystems, e.g. $\text{W}(\text{S}_{1-x}\text{Se}_x)_2$ and $(\text{Mo}_x\text{Ta}_{1-x})\text{Te}_2$. Each monolayer was constructed with mirror-plane symmetry, a lattice parameter of $\SI{3.18}{\AA}$, and a thickness of $\SI{3.19}{\AA}$, as per ASE's convention. To avoid surface-surface interactions across the periodic boundaries, each monolayer was prepared with a $\SI{40.0}{\AA}$ vacuum.

All DFT calculations were performed using the Vienna Ab initio Simulation Package (VASP) \cite{kresse1993ab, kresse1996efficiency, kresse1996software} with the projector-augmented-wave (PAW) \cite{kresse1999ultrasoft} method and the Perdew–Burke–Ernzerhof (PBE) exchange-correlation functional \cite{perdew1996generalized} on Clemson University's Palmetto Cluster \cite{antao2024modernizing}. A plane-wave kinetic-energy cutoff of $\SI{500}{eV}$ was employed. Brillouin-zone integrations were carried out using a $\Gamma$-centered ($3\times 3\times 1$) Monkhorst–Pack $k$-point mesh \cite{monkhorst1976special}. Structural relaxations were performed using a Gaussian smearing scheme with $\sigma = \SI{0.10}{eV}$, with a maximum of 400 ionic steps and up to 100 electronic iterations per ionic step. The lattice parameters were relaxed in-plane while the out-of-plane lattice parameter was constrained, corresponding to fixed and variable lattice degrees of freedom along the relevant directions.

After the relaxation above, also using VASP, we carried out molecular dynamics with a Langevin thermostat and a Parrinello-Rahman barostat \cite{allen2017computer, parrinello1980crystal, parrinello1981polymorphic}, with a timestep of $\SI{1}{fs}$ for 5,000 steps, starting at $\SI{50}{K}$ and ending at $\SI{400}{K}$ with a fixed pressure of $\SI{0}{kbar}$. Langevin friction coefficients for each species, as well as for the simulation cell, were set to $\SI{10}{ps^{-1}} = \SI{10}{THz}$, with a fictitious cell mass of $\SI{1000}{amu}$. As done in the structural relaxation, the out-of-plane lattice parameter was fixed. This dataset, involving energies and forces in all snapshots (including relaxation and dynamics), is available on GitHub \cite{he-tmd-dataset}.

We then finetune the MACE-MP-0 large model \cite{batatia2025foundation, tompa2026fine} using rank-$4$ low rank adaptation (LoRA) \cite{hu2021lora} and each of the dataset subsets below, creating two different potentials:

\begin{itemize}
    \item The TCE-selected subset $\mathcal{X}_\text{TCE}$
    \item The MACE-selected subset $\mathcal{X}_\text{MACE}$
\end{itemize}

For validation, we draw 5 random trajectory pairs from the remaining dataset $\mathcal{X}_r$:

\begin{equation}
    \begin{aligned}
        \mathcal{X}_r &= \mathcal{X}\setminus\left(\mathcal{X}_\text{TCE}\cup\mathcal{X}_\text{MACE}\right) \\
        \mathcal{X}_\text{val} &\subset \mathcal{X}_r
    \end{aligned}
\end{equation}

and benchmark each potential against 10 trajectory pairs from the remaining dataset, i.e. $\mathcal{X}_r\setminus\mathcal{X}_\text{val}$.

Using the finetuned potentials, we compute the vibrational spectra for the $M(X,X')_2$ monolayers, where $M\in\{\text{W}, \text{Mo}, \text{Ta}\}$ and $\{X,X'\}\in\{\{\text{S},\text{Te}\}, \{\text{Se},\text{Te}\}\}$. We choose these particular $M(X,X')_2$ subsystems so that we can reasonably recover an order-disorder transition from Monte Carlo (MD) simulations without finely resolving low temperatures, as well as additionally focusing on one structural phase, i.e. the $H$ phase. Namely, for $M\in\{\text{W},\text{Mo}\}$, it was predicted using a cluster expansion model that the $M(X,X')_2$ subsystems have $H$ symmetry and an order-disorder transition temperatures (ODTTs) above $\SI{100}{K}$ \cite{linderalv2022high}, in contrast to subsystems such as $\text{W}(\text{S},\text{Se})_2$ with an ODTT of $\SI{30}{K}$, or $(\text{Mo},\text{W})\text{Te}_2$ with an ODTT of $\SI{10}{K}$. 

Importantly, we do not assume that each monolayer system is fully random, i.e. there exists some residual chemical order. In principle, one could use hybrid MC/MD routines with any interatomic potential to sample chemical ordering \cite{jeffries2025prediction}. However, for systems with significant structural relaxations upon swapping two particles, this routine is intractable, since this necessitates relaxing before accepting or rejecting an MC swap. This is exaggerated even more by MLIP evaluations, which are significantly expensive relative to classical potentials. On-lattice cluster expansion approaches fitted to relaxed configurations are ideal for this problem, in that they can accurately reproduce relaxed DFT energies without explicit structural relaxation \cite{shapeev2017accurate, kostiuchenko2019impact, vazquez2024deciphering}.

As such, we fit a TCE model (with the same feature set from Fig.~\ref{fig:ws2-clusters}) using the same structural relaxation data $\mathcal{X}$, and use that model to run canonical MC simulations at different temperatures to generate partially chemically-ordered samples. We provide more details for these MC simulations in Sec.~\ref{sec:tce+mc}.

Then, for each sample chemically equilibrated at a 
temperature $T$ (with temperatures 100, 200, 300, and 400 K), we compute the vibrational spectrum by running molecular dynamics using Torch-Sim \cite{paszke2019pytorch, cohen2025torchsim}, with the following sequence:

\begin{itemize}
    \item Structural relaxation at $\SI{0}{K}$ using FIRE \cite{bitzek2006structural}, with variable lattice parameters using the Frechet cell filter
    \item NVT equilibration at a temperature $T$ using the Langevin integrator \cite{allen2017computer, leimkuhler2016efficient} with a timestep $\SI{1}{fs}$ and a friction coefficient $\SI{10}{ps^{-1}}$ for 100,000 steps
    \item NVE run for 100,000 steps
\end{itemize}

During the NVE run, we compute the mass-weighted velocity autocorrelation function (VACF) and the resulting vibrational density of states (VDOS) via the Fourier transform of the VDOS \cite{frenkel1997understanding}:

\begin{equation}\label{eq:vacf-and-vdos}
    \begin{aligned}
        C(t) &\propto \sum_i m_i\left\langle \mathbf{v}_i(0)\cdot\mathbf{v}_i(t)\right\rangle \\
        g(\omega) & \propto \int_0^\infty C(t) \exp\left(\iu\omega t\right)\;\mathrm{d}t
    \end{aligned}
\end{equation}

Here, we compute the VACF at constant energy rather than constant temperature to avoid known artifacts of a relatively strongly coupled Langevin thermostat on vibrational dampening \cite{heinz2026thermostats}. Averages above are computed within Torch-Sim by sampling every 10 timesteps and averaged over a moving window of 2,000 samples, corresponding to a spectral resolution of approximately $(c\times 2000\times 10\times \SI{1}{fs})^{-1} \approx \SI{1.668}{cm^{-1}}$, where $c = \SI{3.0e-5}{cm/fs}$ is the speed of light.

%% file: sections/results.tex
\section{Results}

\begin{figure}[H]
    \centering
    \includegraphics[width=\linewidth]{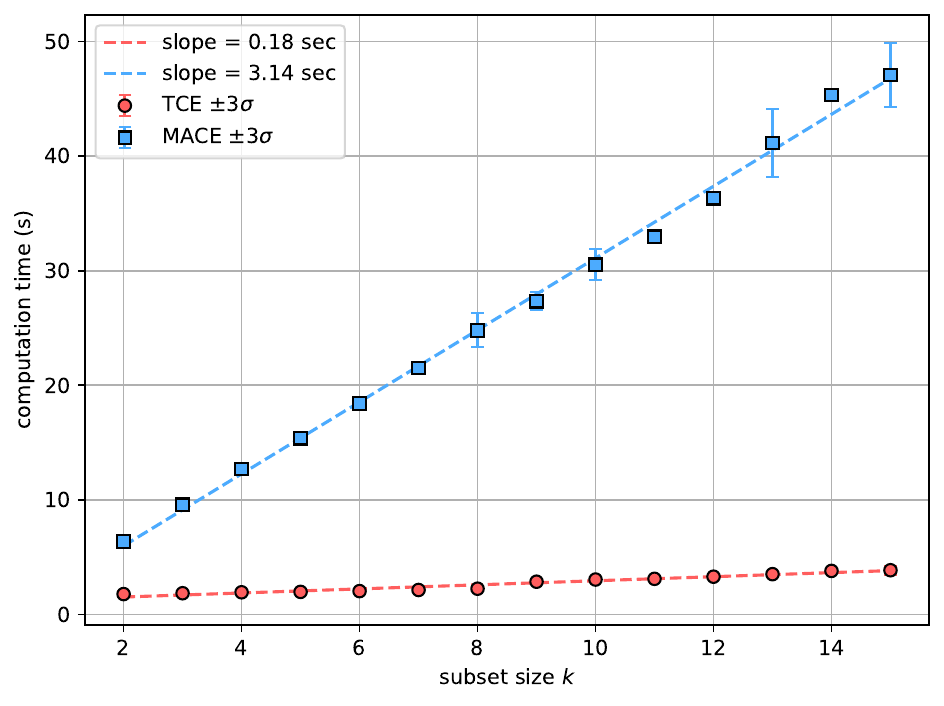}
    \caption{Computation time (with a sample size of $5$) to compute TCE features or MACE features of a size-$k$ subset of $4\times 4$ TMD monolayers on a single Intel Xeon CPU E5-2660  $\SI{2.20}{GHz}$ CPU.}
    \label{fig:timing-benchmark}
\end{figure}

In regard to the computational cost, we find that the computation time to compute the data matrix corresponding to TCE features is significantly smaller than that of MACE features. Namely, the computation time of a TCE feature vector is about $\SI{3.14}{sec} / \SI{0.18}{sec} \approx 17\text{x}$ faster than a MACE feature vector on a single CPU (Fig.~\ref{fig:timing-benchmark}). The main reason for this is that the tensor contractions to construct TCE features (Eq.~\eqref{eq:tensor-contraction}) are simpler mathematical operations than constructing MACE descriptors. Additionally, the TCE features describe a fixed, on-lattice system, so the neighbor list does not need to be recomputed, meaning that the $\mathbf{A}^{[\ell]}$ tensors need to only be computed once if all samples share the same lattice positions. We note that in order to generate a fair comparison in Fig.~\ref{fig:timing-benchmark}, each subset has its own independent cache, i.e. each TCE data point independently computes $\mathbf{A}$.

A standard workflow would typically perform the MACE calculations on a GPU, rather than a single CPU, narrowing the performance gap relative to TCE calculations on a single CPU. However, the tensor contractions performed by \verb|tce-lib| are well suited to GPU acceleration, especially using libraries that support sparse tensors like PyTorch \cite{paszke2019pytorch}. We save this implementation for a future work. For now, though, the advantage is clear, in that we can efficiently perform subset selection in a resource-constrained environment.

\begin{figure}[H]
    \centering
    \includegraphics[width=\linewidth]{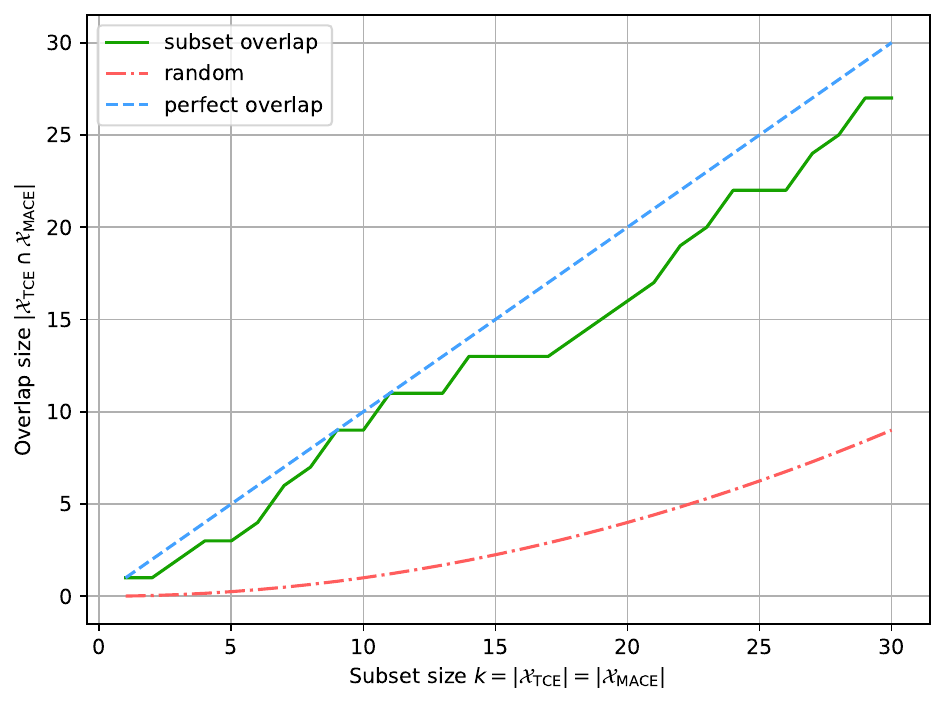}
    \caption{Consistency index $I = |\mathcal{X}_\text{TCE}\cap \mathcal{X}_\text{MACE}|$ vs. subset size $k = |\mathcal{X}_\text{TCE}| = |\mathcal{X}_\text{MACE}|$ bounded by the perfect case $I = k$ and the random case $I = k^2/n$.}
    \label{fig:consistency-v-subset}
\end{figure}

We find that the overlap between the two chosen subsets significantly exceeds that of a random overlap, meaning that the TCE subset selection is well-selecting the basis used to construct MACE features (Fig.~\ref{fig:consistency-v-subset}).

\begin{figure}[H]
    \centering
    \includegraphics[width=\linewidth]{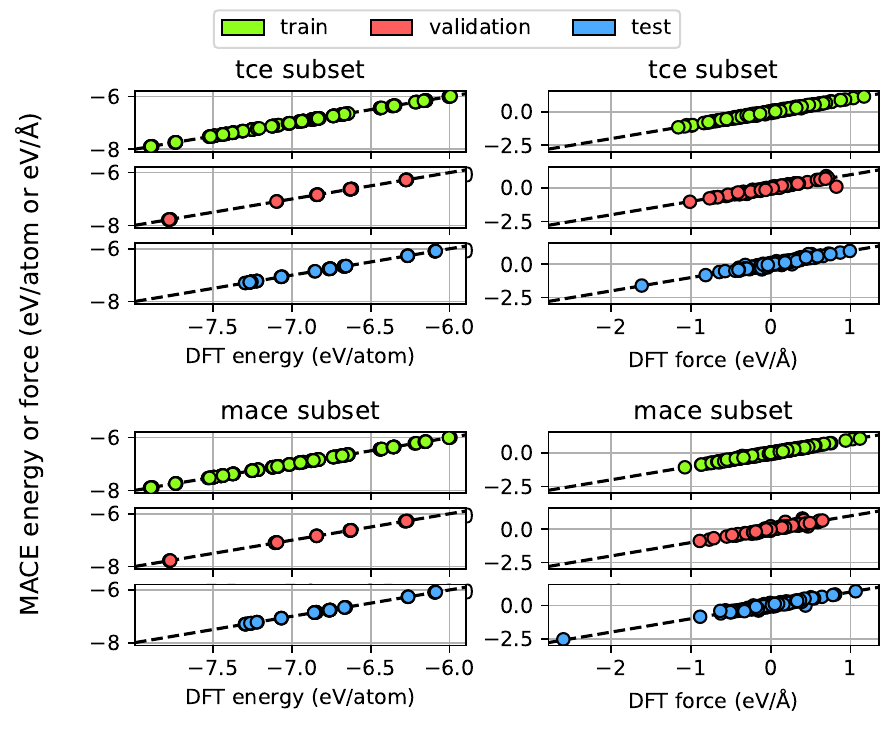}
    \caption{Parity plots of energies (left) and forces (right) for the potentials fine-tuned on the TCE- (top) and MACE-selected (right) subsets. Each panel respectively contains $0.5\%$ and $0.01\%$ of the full energies and forces data for clearer visualization.}
    \label{fig:parity}
\end{figure}

We additionally find that both potentials reasonably predict both the DFT energies and forces in their respective training sets, as well as the shared validation and test sets (Fig.~\ref{fig:parity}).

\begin{figure}[H]
    \centering
    \includegraphics[width=\linewidth]{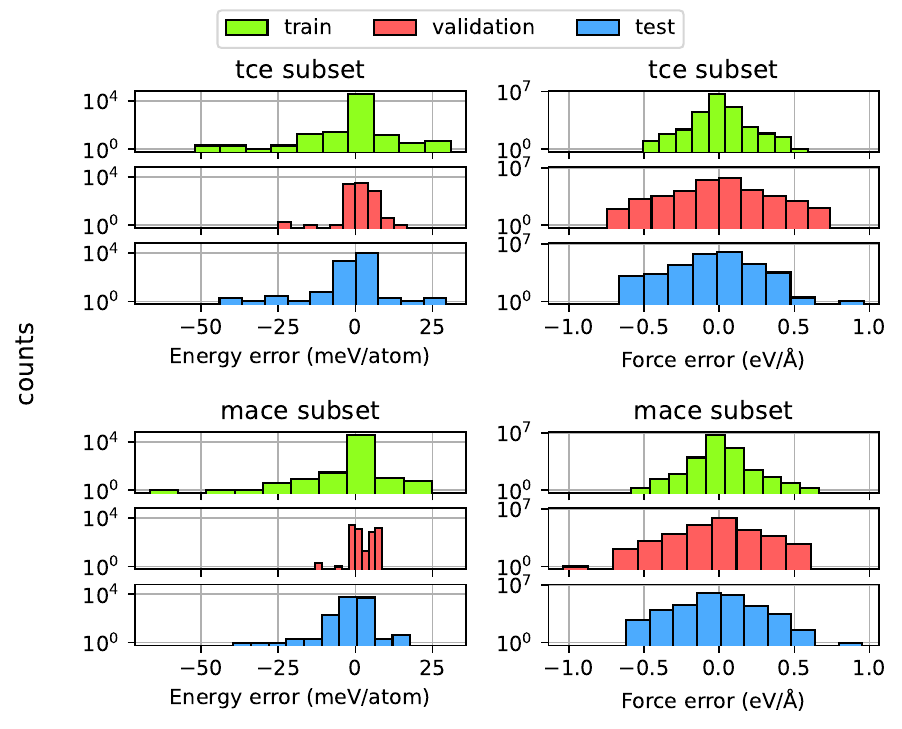}
    \caption{Distribution of residuals ($\text{predicted} - \text{DFT}$) using the TCE and MACE selected subsets (top and bottom respectively) for both energies (left) and forces (right). Here, counts on the vertical axis are log-scaled.}
    \label{fig:residuals}
\end{figure}

Each distribution of residuals is relatively symmetric about $0$, with respective widths on the order of $\SI{10}{meV/atom}$ and $\SI{0.5}{eV/\AA}$ on all dataset splits, indicating further that the potentials are not overfitted (Fig.~\ref{fig:residuals}).

\begin{figure}[H]
    \centering
    \begin{subfigure}[t]{0.45\linewidth}
        \centering
        \includegraphics[width=0.45\linewidth]{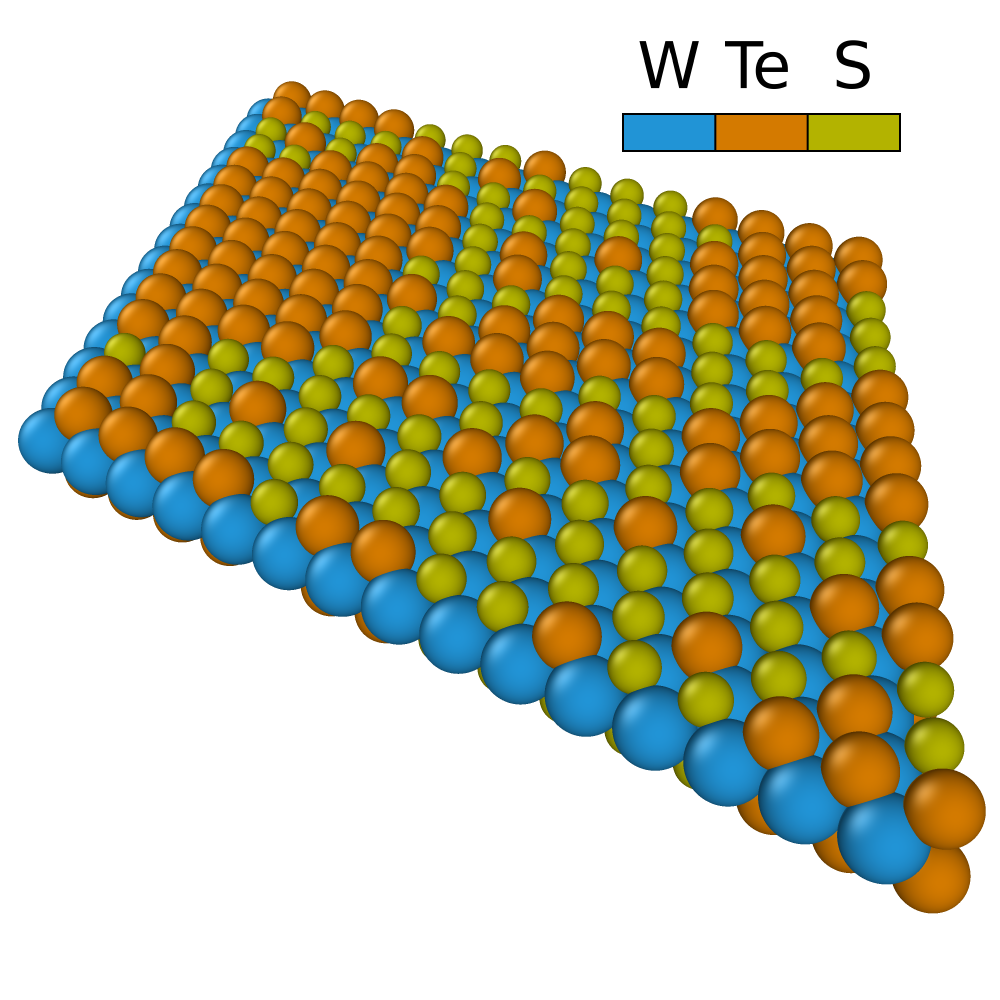}
        \includegraphics[width=0.45\linewidth]{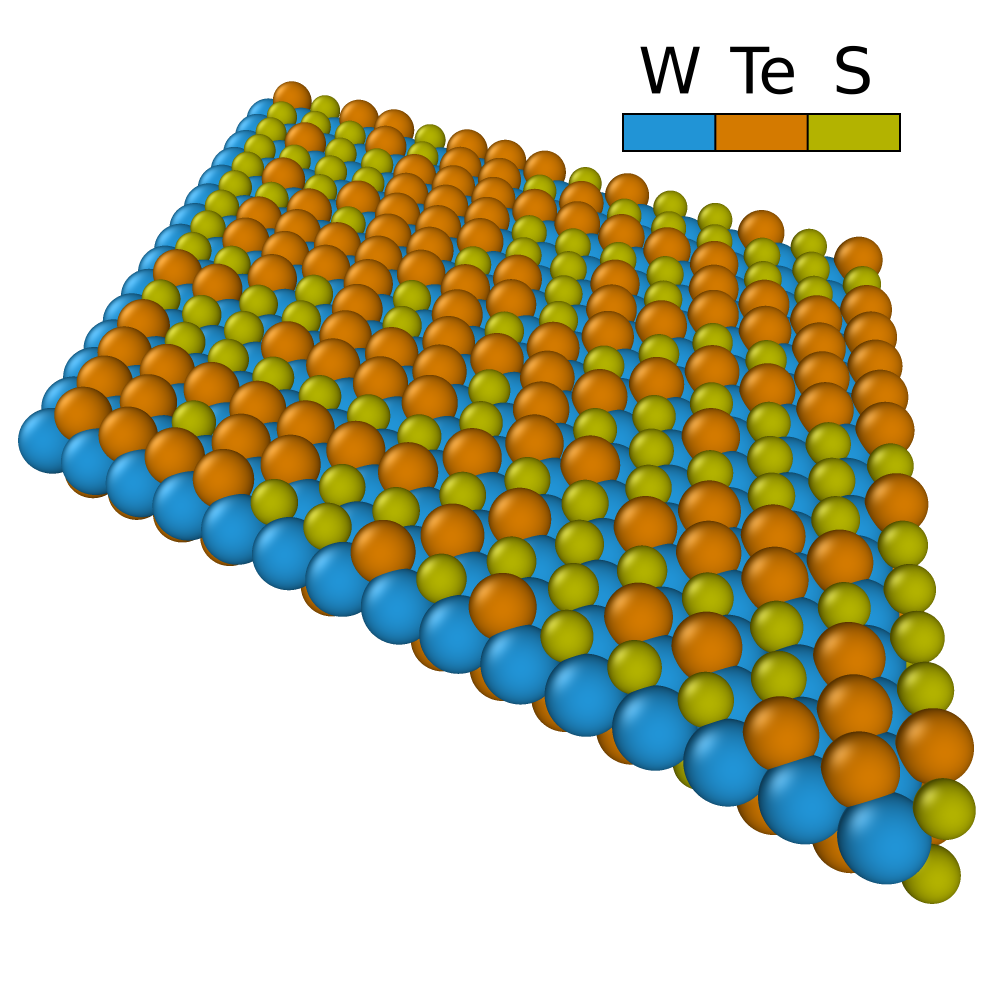}
        \caption{\small $\text{W}(\text{S},\text{Te})_2$}
        \label{fig:WSTe}
    \end{subfigure}
    \begin{subfigure}[t]{0.45\linewidth}
        \centering
        \includegraphics[width=0.45\linewidth]{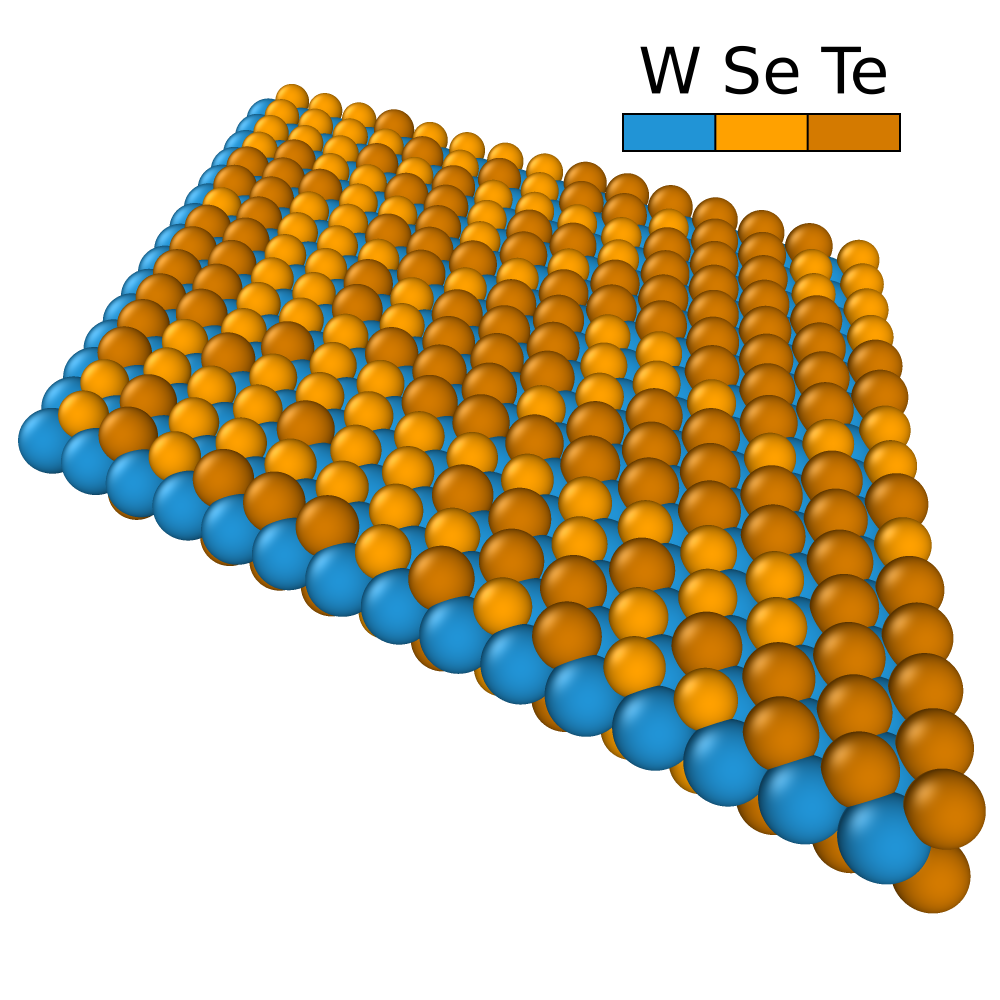}
        \includegraphics[width=0.45\linewidth]{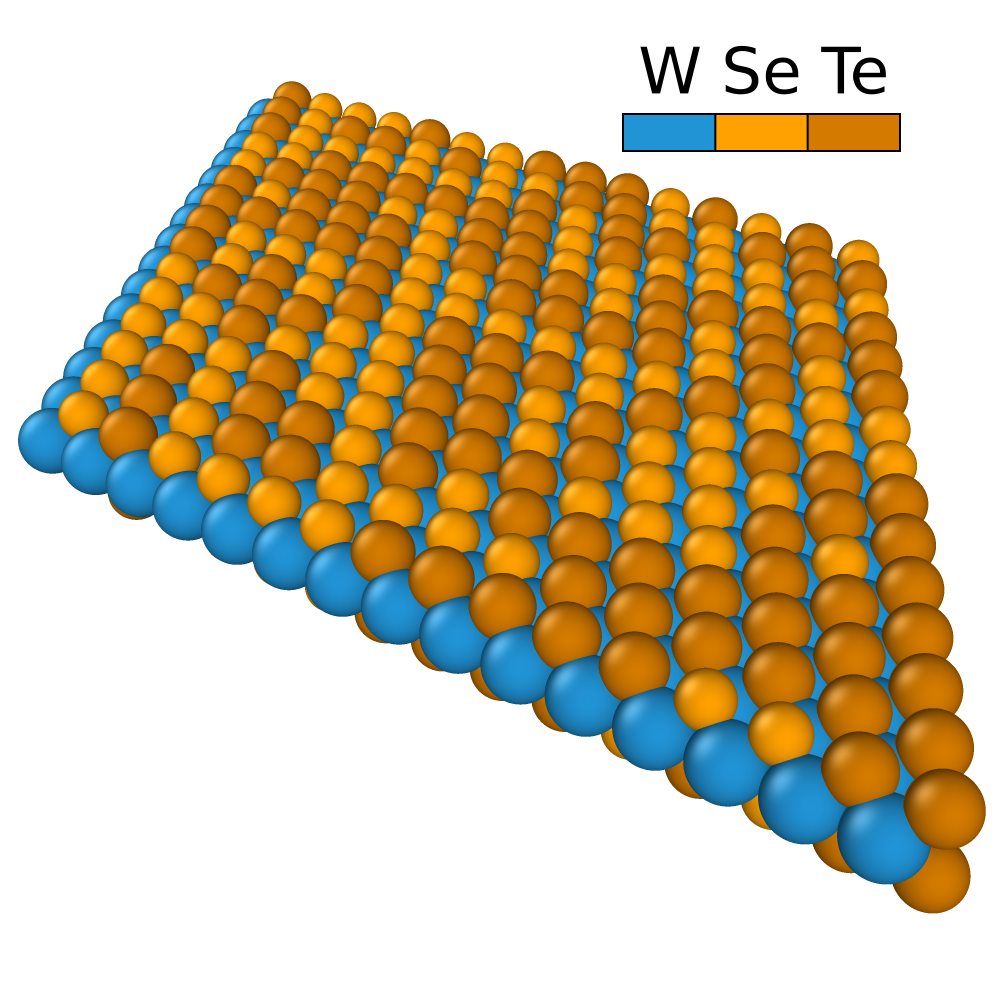}
        \caption{\small $\text{W}(\text{Se},\text{Te})_2$}
        \label{fig:WSeTe}
    \end{subfigure}
    \begin{subfigure}[t]{0.45\linewidth}
        \centering
        \includegraphics[width=0.45\linewidth]{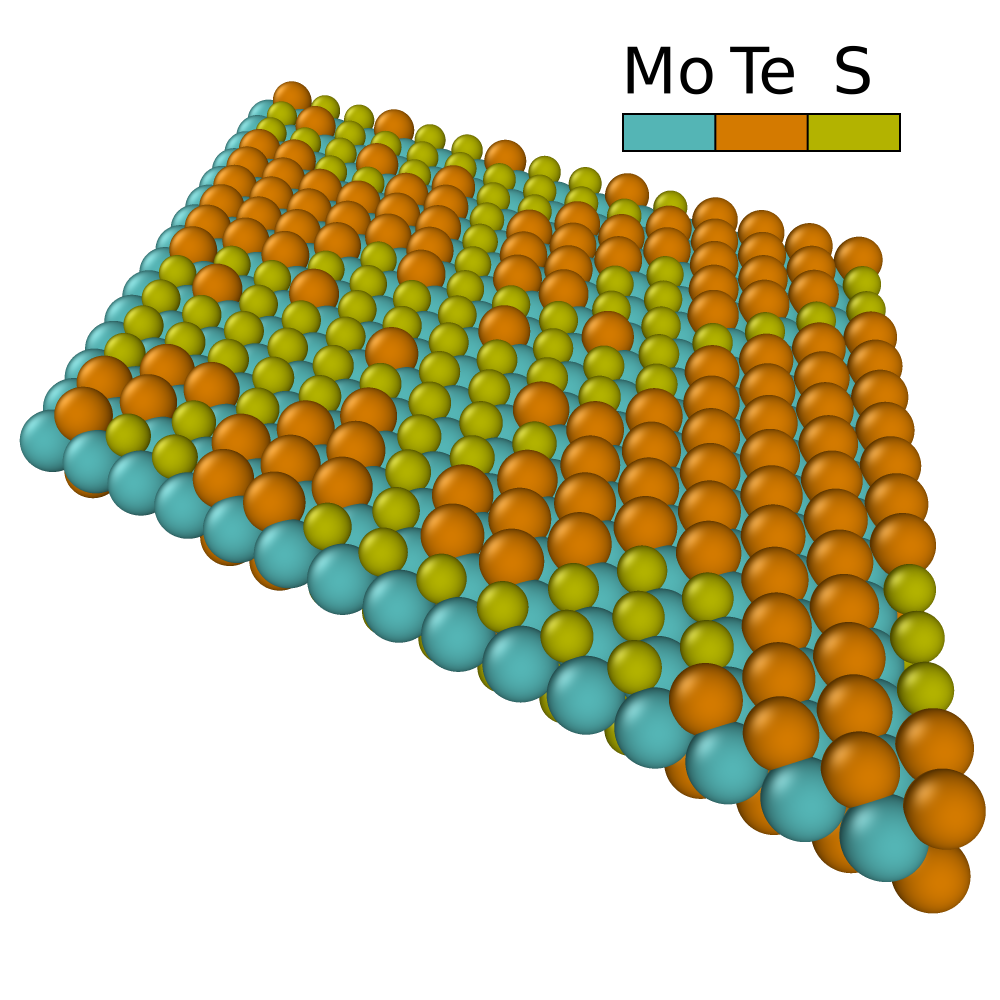}
        \includegraphics[width=0.45\linewidth]{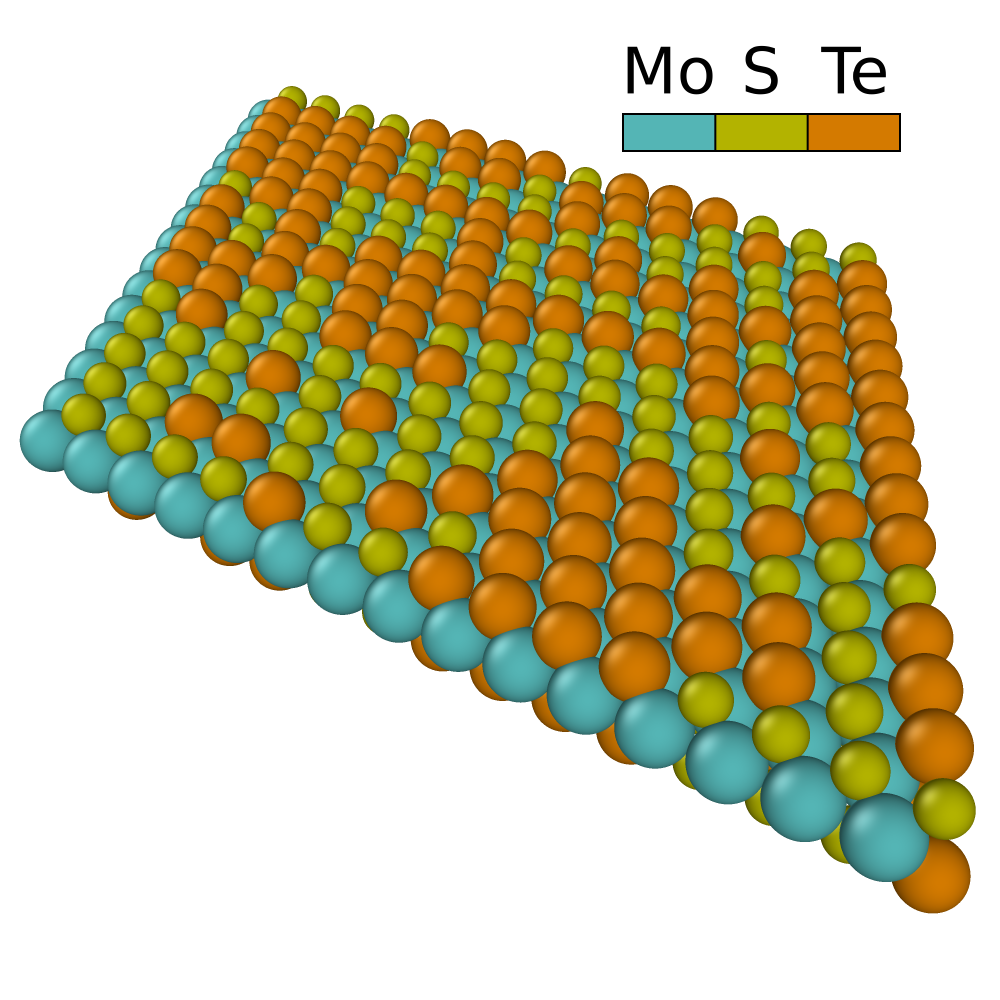}
        \caption{\small $\text{Mo}(\text{S},\text{Te})_2$}
        \label{fig:MoSTe}
    \end{subfigure}
    \begin{subfigure}[t]{0.45\linewidth}
        \centering
        \includegraphics[width=0.45\linewidth]{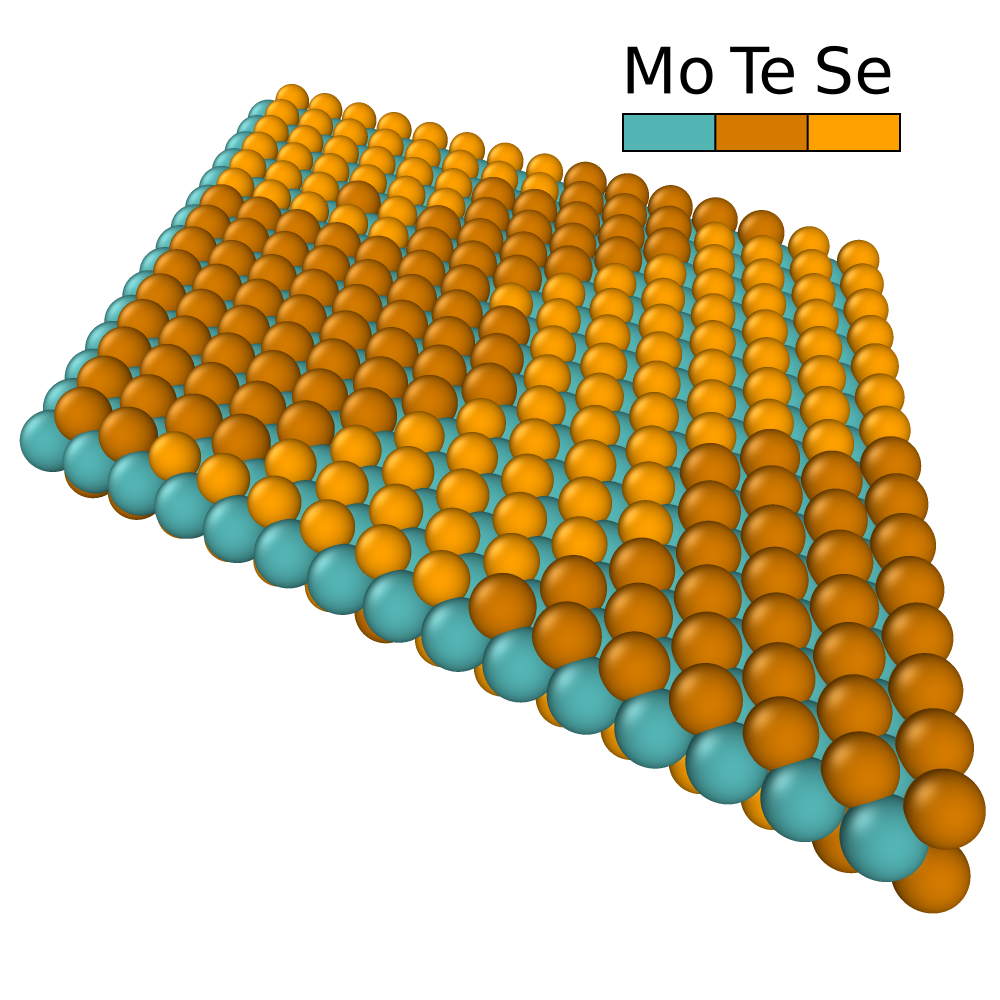}
        \includegraphics[width=0.45\linewidth]{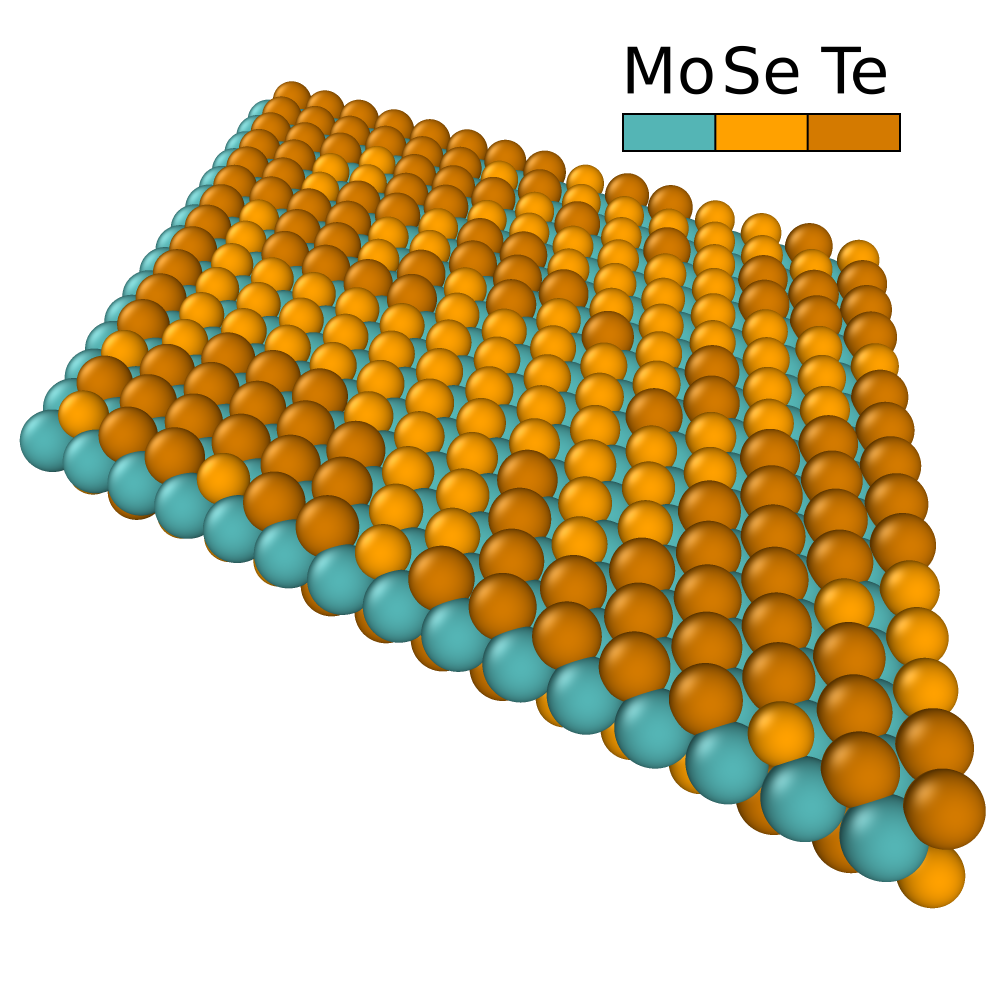}
        \caption{\small $\text{Mo}(\text{Se},\text{Te})_2$}
        \label{fig:MoSeTe}
    \end{subfigure}
    \begin{subfigure}[t]{0.45\linewidth}
        \centering
        \includegraphics[width=0.45\linewidth]{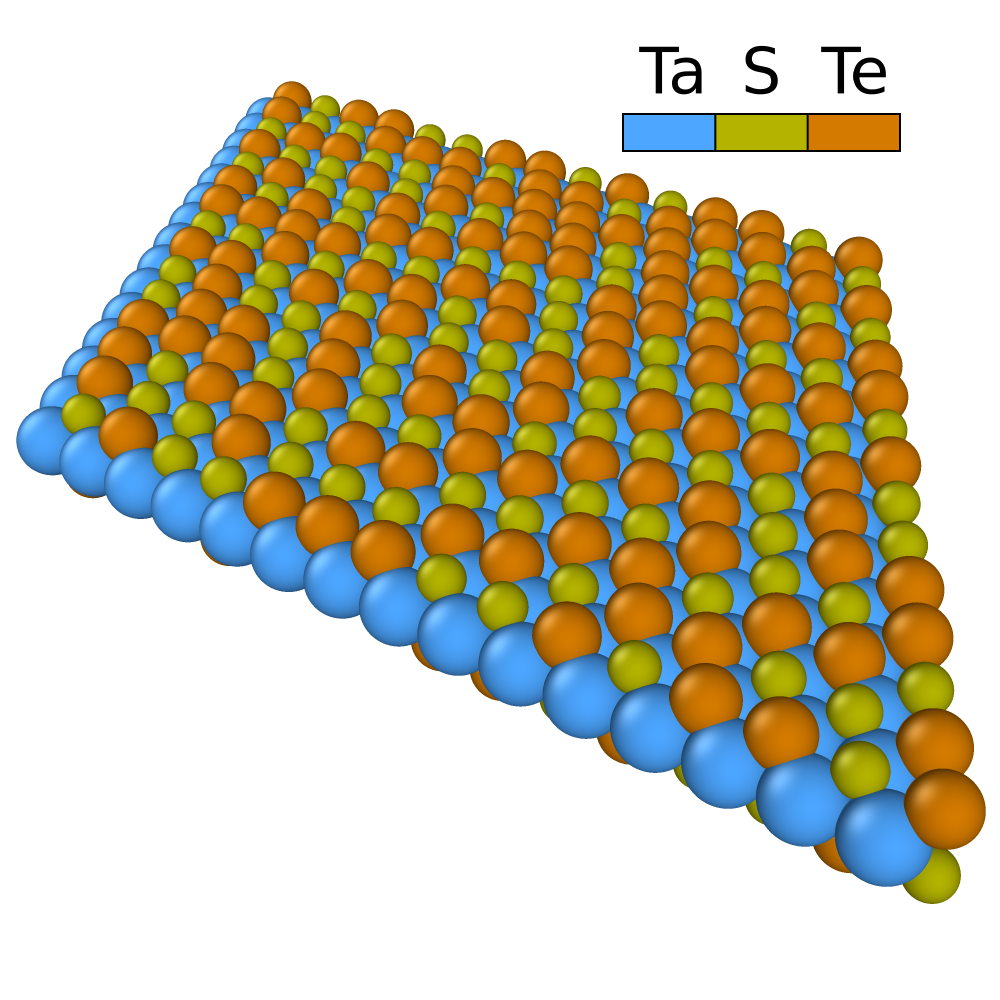}
        \includegraphics[width=0.45\linewidth]{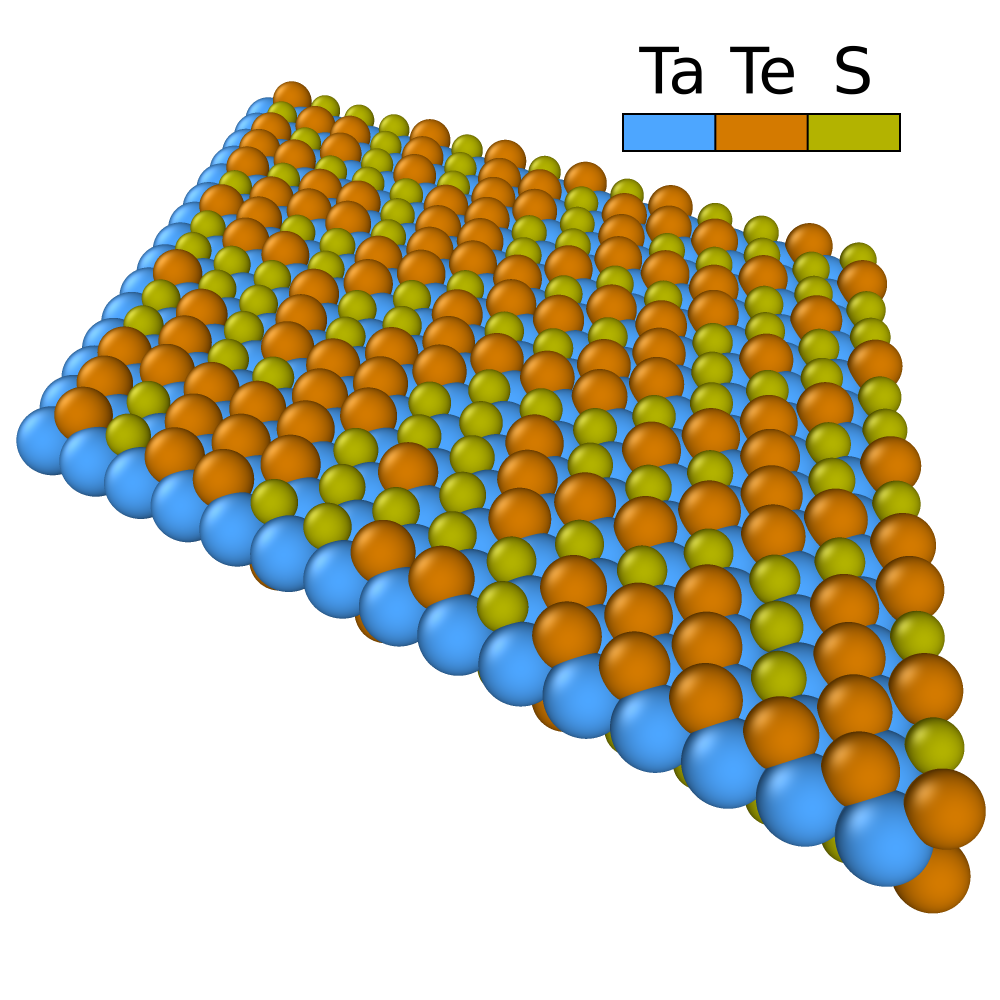}
        \caption{\small $\text{Ta}(\text{S},\text{Te})_2$}
        \label{fig:TaSTe}
    \end{subfigure}
    \begin{subfigure}[t]{0.45\linewidth}
        \centering
        \includegraphics[width=0.45\linewidth]{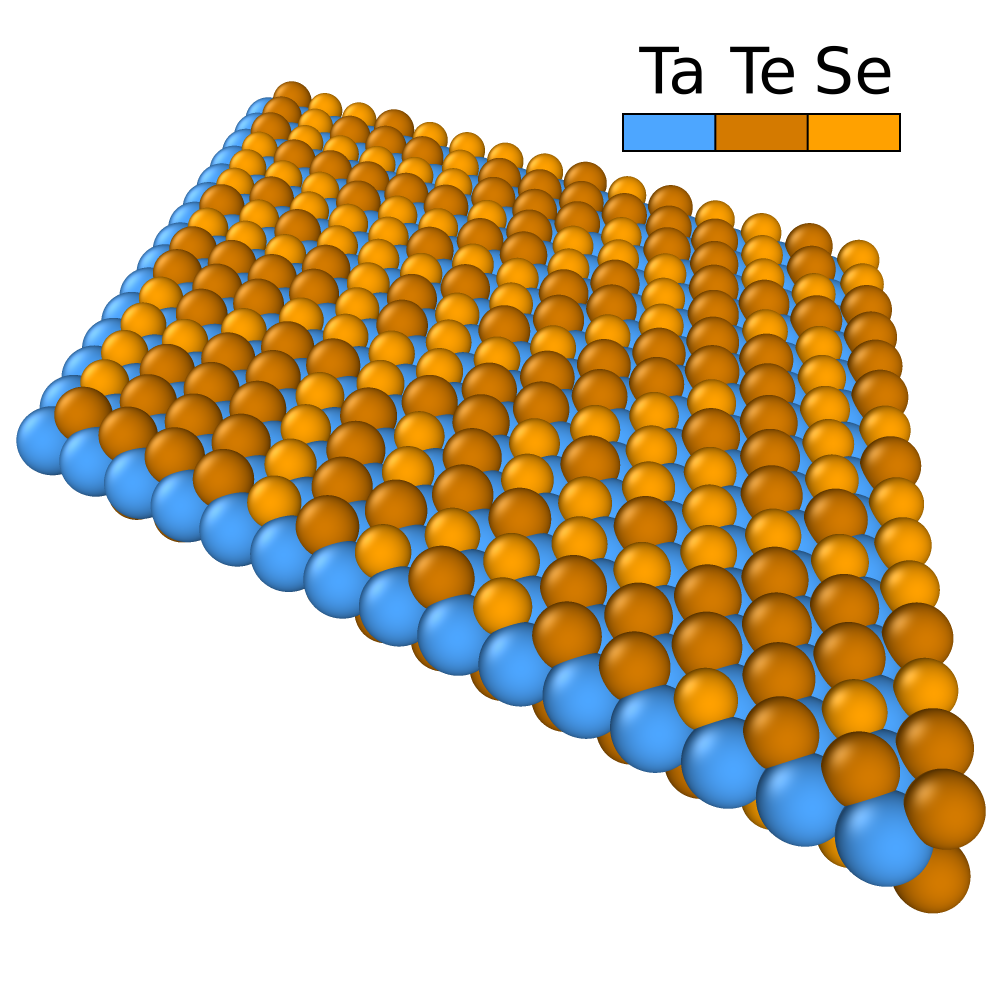}
        \includegraphics[width=0.45\linewidth]{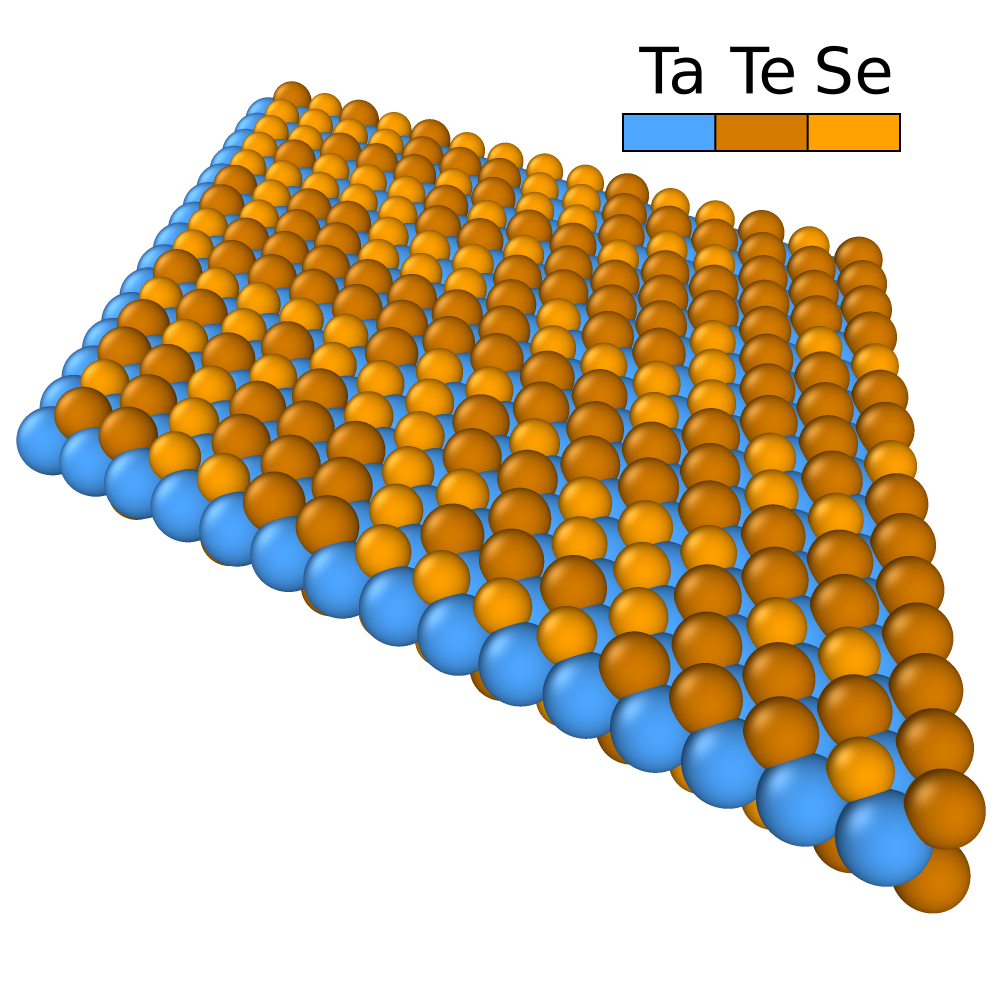}
        \caption{\small $\text{Ta}(\text{Se},\text{Te})_2$}
        \label{fig:TaSeTe}
    \end{subfigure}
    \caption{Extreme $M(X,X')_2$ configurations visualized with OVITO \cite{stukowski2010visualization} at $\SI{100}{K}$ (left subpanels) and at $\SI{400}{K}$ (right subpanels).}
    \label{fig:visualization}
\end{figure}

We find that, at low temperatures, the TCE model predicts substantial $\text{Te}$ clustering in $\text{W}(\text{S},\text{Te})_2$, $\text{Mo}(\text{S},\text{Te})_2$, and $\text{Mo}(\text{Se},\text{Te})_2$. This chemical ordering, although suppressed relative to $\SI{100}{K}$, survives at relatively high temperatures (i.e. $\SI{400}{K}$) in these systems. (Fig.~\ref{fig:visualization}). In the other TMD monolayers, we find more moderate low-temperature chemical ordering.

\begin{figure}[H]
    \centering
    \includegraphics[width=\linewidth]{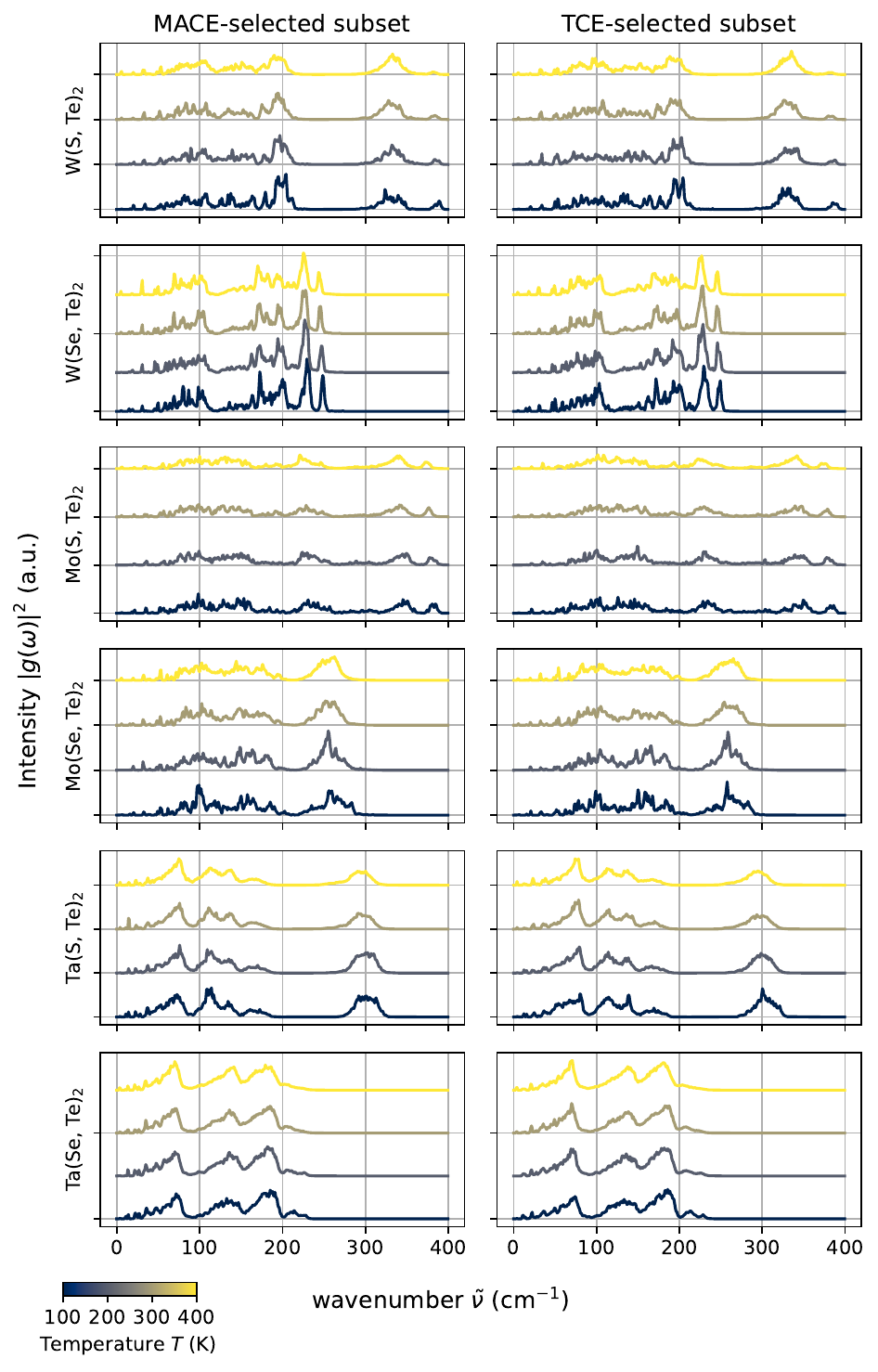}
    \caption{VDOS intensities $|g(\omega)|^2$ from the two MACE potentials finetuned on $\mathcal{X}_\text{MACE}$ (left) and $\mathcal{X}_\text{TCE}$ (right). Each row represents a different $M(X,X')_2$ system, with each color representing an MD run at a given temperature.}
    \label{fig:vdos}
\end{figure}

Using the two MACE potentials, we then compute the VDOSs at varying temperatures by Fourier transforming the VACFs. Here, we plot the result as a function of wavenumber $\tilde{\nu} = 2\pi\omega/c$. We additionally show the raw VACF data in Sec.~\ref{sec:vacf}. For both potentials, the VDOSs are qualitatively reasonable. For example, the monolayers containing S rather than Se have higher frequency peaks, with peaks likely corresponding to S's smaller weight relative to Se. Furthermore, at higher temperatures, we find that some peaks broaden, e.g. the $\SI{300}{cm^{-1}}$ peak in $\text{Ta}(\text{S},\text{Te})_2$ (Fig.~\ref{fig:vdos}), which is notably well-mixed even at low temperatures (Fig.~\ref{fig:TaSTe}).

However, most importantly, both sets of VDOSs are very similar for each $M(X,X')_2$ system, with differences only at very small spectral resolutions. For example, in $\text{Ta}(\text{S},\text{Te})_2$, the low-temperature peak around $\SI{300}{cm^{-1}}$ is marginally broader. The important feature of the VDOSs, though, is that the peak locations are invariant with respect to the subset chosen to finetune the potential. This is unsurprising, given that the overlap of $\mathcal{X}_\text{TCE}$ and $\mathcal{X}_\text{MACE}$ is quite large (Fig.~\ref{fig:consistency-v-subset}). Thus, this showcases that we can successfully replace a significant portion of compute power, i.e. the computation of atomic descriptors during the selection of a training set, with a significantly more cost-efficient surrogate model. We expect that this result extends to the prediction of a much larger class of dynamical properties in concentrated alloys, e.g. dynamical structure factors \cite{he2025resolving}, phonon lifetimes \cite{dong2024thermal}, lattice thermal conductivity \cite{conley2025heat}, etc., but we save this exploration for future work.

%% file: sections/discussion.tex
\section{Discussion}

We show, in this work, that TCE can be used as a cost-effective surrogate for the chemical complexity captured within MACE/ACE descriptors. This is particularly relevant for the high-throughput screening of materials, in which a significant bottleneck is the selection of samples with diverse chemistry to perform \textit{ab initio} calculations on.

This is particularly relevant in the setting of building autonomous labs for the discovery and development of materials \cite{stach2021autonomous}. In this setting, it is advantageous to inject theory-in-the-loop, in the sense that the workflow proposes a class of materials living within some thermodynamic space, generates an MLIP to accurately describe that region of thermodynamic space, and performs the desired computations using the MLIP, often MD. prior to training said MLIP, one needs to choose a sufficiently chemically diverse subspace of configurations within that region of thermodynamic space.

In this work, we show that the popular ACE-class of potentials can be sufficiently pre-screened using computationally cheaper surrogate TCE features. This is principally advantageous for two reasons. More obviously, this pre-screening is, theoretically, more efficient on equivalent hardware, and empirically more efficient on single CPUs.

Furthermore, the TCE cluster counts defined in Eq.~\eqref{eq:tensor-contraction} are well-suited for basic hardware, including single CPUs. From the numerical experiments in this work, we find that each TCE feature vector takes on the order of $\SI{0.1}{sec}$ on a single CPU, meaning that it only takes roughly a day to compute feature vectors for 100,000 configurations without parallelization. This matters in a very practical scenario where, for example, an autonomous lab has to wait in an HPC queue for specialized hardware. Then, if the autonomous lab uses TCE surrogates in the pre-screening step, the agent can avoid a potentially long GPU queue.

Similarly, this work potentially opens the route for \textit{ab initio} accurate MC/MD simulations in hardware-constrained environments. For example, TCE can be used as a surrogate Hamiltonian to resolve configurational energy/entropy and an MLIP like MACE to resolve vibrational energy/entropy, effectively using a hybrid TCE/MACE routine as a more computationally efficient thermodynamic interatomic potential, similar to recent works on on-lattice harmonic potentials \cite{lee2024machine, lee2025accelerating, tolborg2023low}.

Moreover, there are many potential routes for optimization of the feature vectors. The more obvious route is to implement TCE within PyTorch, as previously described. Moreover, a potentially advantageous route is to build a workflow that does not generate configurations independently, and instead iteratively generates chemically diverse configurations from prior ones. In this scenario, one can compute the truncated $\Delta\mathbf{N}$ from Eq.~\eqref{eq:swap-shortcut}, whose computation time is independent of configuration size.

%% file: sections/limitations.tex
%\section{Limitations}

The primary limitation in this work is creating a dataset that includes defects, which are of particular interest to the TMD community. For point-like defects that sit within the same lattice as the atom types, this is reasonably amenable by adding a vacancy type in the tensor contractions in Eq.~\eqref{eq:tensor-contraction}, i.e. by treating the vacancy as its own alloying type. Representing other types of defects, such as interstitials or ad-atoms, is not as trivial. A potential route is to augment the lattice to include extra lattice sites that can capture these off-lattice defects, similar to prior work done for using the cluster expansion technique to model vacancy diffusion in concentrated alloys, in which the transition state for a vacancy hop is represented by points in a superlattice \cite{jeffries2026first, li2021predicting}.

A secondary limitation is that TCE restricts its internal representation to rotationally invariant features and therefore does not retain equivariant vector- or tensor-valued information. This restriction is sufficient for predicting scalar quantities such as the total potential energy, and equivariant forces can still be obtained by differentiating the energy with respect to atomic positions. However, it may limit the model’s ability to represent interactions whose natural degrees of freedom are directional or tensorial, particularly when those degrees of freedom must be propagated through multiple layers of the representation. Equivariant architectures such as NequIP and MACE address this issue by maintaining features that transform according to nontrivial representations of the rotation group \cite{batzner20223, batatia2022mace}.

This limitation may be important for potentials describing magnetic materials, including ferritic systems. Magnetic moments are vector-like degrees of freedom, and their coupling to atomic structure can involve changes in both moment magnitude and orientation. In iron, magnetic contributions have been associated with phase stability, elastic behavior near the Curie temperature, defect energetics, and structural transformations \cite{chapman2022machine}. A TCE model trained only on invariant structural features may therefore require substantially higher body order or additional scalar contractions to capture such effects accurately, and it cannot directly provide equivariant magnetic observables. Extending TCE to include equivariant channels so that the spanning relation in Eq.~\eqref{eq:spanning-relation} has some equivariant expressiveness would be a natural direction for future work.

%% file: sections/conclusions.tex
\section{Conclusions}

In this work, we prove that TCE features span an invariant subspace of MACE features. We then show that, in some model alloyed TMD systems, TCE features act as efficient, surrogate pre-screeners for MACE features, by means of performing maximum-entropy subset selection on a high-entropy monolayer TMD dataset. We then compare maximum-entropy subsets selected by both TCE and MACE features to finetune two foundational MACE potentials to calculate VDOSs on some alloyed TMD systems, showing that the two potentials predict nearly identical VDOSs in all of the model systems at all tested temperatures. This approach advances the broader goal of tractable predictions of alloy thermodynamics in concentrated alloys, further opening the path towards the efficient sampling of chemical diversity at-scale. Furthermore, our results establish a direct connection between on-lattice-based descriptions of chemical diversity and representations used in off-lattice MLIPs, providing a foundation for using well-established on-lattice methods to inform the study of chemically complex materials with off-lattice MLIPs.

%% file: sections/data_availability.tex
\section{Data Availability}

The full TMD dataset used to train the MLIPs and TCE model in this work, as well as the full scripts to reproduce our results, are available on GitHub \cite{he-tmd-dataset, finetuning, ordered-tmd-tce}.

%% file: sections/acknowledgements.tex
\section{Acknowledgments}

We thank Panchapakesan Ganesh for insightful discussion that contributed to this work.

The authors acknowledge support by the U.S. Department of Energy, Office of Science, Office of Workforce Development for Teachers and Scientists, Office of Science Graduate Student Research (SCGSR) program. The SCGSR program is administered by the Oak Ridge Institute for Science and Education for the DOE under contract number DE‐SC0014664.

J. Jeffries and E. Martinez acknowledge support from the U.S. Department of Energy, Office of Basic Energy Sciences, Materials Science and Engineering Division under Award No. DE-SC0022980.

E. Zarkadoula was supported by the Center for Nanophase Materials Sciences, (CNMS), which is a US Department of Energy, Office of Science User Facility at Oak Ridge National Laboratory.

Additionally, this material is based on work supported by the National Science Foundation under Grant Nos. MRI\# 2024205, MRI\# 1725573, and CRI\# 2010270 for allotment of compute time on the Clemson University Palmetto Cluster.

%% file: sections/disclaimer.tex
\section{Disclaimer}

Any opinions, findings, and conclusions or recommendations expressed in this material are those of the author(s) and do not necessarily reflect the views of the National Science Foundation.

%% file: sections/appendix.tex
\appendix

\section{Appendix}

\subsection{Outline of proof}\label{sec:proof-outline}

\begin{figure}[H]
    \centering
    \begin{tikzpicture}[
        atom/.style={circle, draw, thick, minimum size=7mm, inner sep=0pt},
        rootatom/.style={atom, fill=blue!25},
        neighboratom/.style={atom, fill=orange!35},
        edgearrow/.style={-Latex, thick},
        labelbox/.style={font=\small, align=center}
    ]
    
    % --- atoms ---
    \node[rootatom] (i) at (0,0) {$i$};
    
    \node[neighboratom] (j1) at (2.0, 0.0) {$j_1$};
    \node[neighboratom] (j2) at (0.0, 2.0) {$j_2$};
    \node[neighboratom] (j3) at (-2.0, 0.0) {$j_3$};
    
    % --- root-neighbor edges ---
    \draw[thick] (i) -- (j1);
    \draw[thick] (i) -- (j2);
    \draw[thick] (i) -- (j3);
    
    % --- neighbor-neighbor edges ---
    \draw[thick, dashed] (j1) -- (j2);
    \draw[thick, dashed] (j2) -- (j3);
    
    % --- set labels ---
    \node[labelbox] (setrn) at (0.0,-1.25) {$\{(i,j_\xi)\;|\; 1\leq\xi\leq \nu\}$ \\ root-neighbor edges};
    \node[labelbox] (setnn) at (0,3.0) {$\{(j_\xi,j_\zeta)\;|\; 1\leq\xi < \zeta \leq \nu\}$ \\ neighbor-neighbor edges};
    
    % --- arrows from set labels to all matching root-neighbor edges ---
    \draw[edgearrow] (setrn.north east) to[out=15,in=270] ($(i)!0.55!(j1)$);
    \draw[edgearrow] (setrn.north) to[out=165,in=200] ($(i)!0.35!(j2)$);
    \draw[edgearrow] (setrn.north west) to[out=165,in=270] ($(i)!0.55!(j3)$);
    
    % --- arrows from set labels to all matching neighbor-neighbor edges ---
    \draw[edgearrow] (setnn.south east) to[out=-15,in=45] ($(j1)!0.55!(j2)$);
    \draw[edgearrow] (setnn.south west) to[out=195,in=135] ($(j2)!0.45!(j3)$);

    \node[labelbox] (setrn) at ($(j1)!0.35!(j2) + (0.2, 0.2)$) {$c_{12}$};
    \node[labelbox] (setrn) at ($(j3)!0.35!(j2) + (-0.2, 0.2)$) {$c_{23}$};

    \node[labelbox] (setrn) at ($(i)!0.5!(j1) + (0.0, 0.25)$) {$a_1$};
    \node[labelbox] (setrn) at ($(i)!0.5!(j2) + (0.25, 0.0)$) {$a_2$};
    \node[labelbox] (setrn) at ($(i)!0.5!(j3) + (0.0, 0.25)$) {$a_3$};
    
    \end{tikzpicture}
    \caption{Root-neighbor vs. neighbor-neighbor edges in an arbitrary $4$-body cluster ($\nu = 3$), respectively corresponding to the adjacency tensors $A_{ij_\xi}^{(a_\xi)}$ and $A_{j_\xi j_\zeta}^{(c_{\xi\zeta})}$, with edge labels $a_\xi$ and $c_{\xi\zeta}$.}
    \label{fig:root-neighbor-v-neighbor-neighbor}
\end{figure}
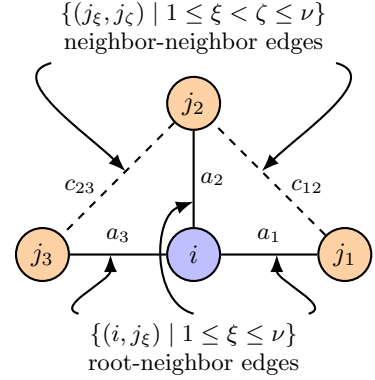

The proof begins by choosing one cluster vertex as the root $i$ and write $f=(\mathbf a,\mathbf c,\boldsymbol\alpha)$, where $a_\xi$ labels the root-neighbor shell, $c_{\xi\zeta}$ labels the neighbor-neighbor shell, and $\sigma_\xi=\alpha_{\xi+1}$. The rooted contribution, i.e. the number of clusters of type $f$ incident to site $i$ is then:

\begin{equation}\label{eq:outline-rooted}
    \begin{aligned}
        \Phi_{i,f} &= X_{i\alpha_1} \sum_{j_1,\ldots,j_\nu} \left[\prod_{\xi=1}^{\nu} A_{ij_\xi}^{(a_\xi)}X_{j_\xi\sigma_\xi}\right]\left[\prod_{\xi<\zeta} A_{j_\xi j_\zeta}^{(c_{\xi\zeta})}\right]
    \end{aligned}
\end{equation}

with the resulting number of clusters of type $f$ being the sum of these root contributions:

\begin{equation}
    N_f = N_{\boldsymbol{\alpha}}^{\ell} = \sum_i \Phi_{i, f}
\end{equation}

where $\ell = \mathbf{a}\oplus\mathbf{c}$ is the topological label of the cluster and $\oplus$ denotes concatenation. We note that Eq.~\eqref{eq:outline-rooted} decomposes the adjacency tensor product into two distinct products, namely that of the root-neighbor edges (labeled with $1\leq\xi\leq\nu$) and that of neighbor-neighbor edges (labeled with $\xi < \zeta$), which is visualized in Figure~\ref{fig:root-neighbor-v-neighbor-neighbor}. MACE features natively capture the root-neighbor edges \cite{batatia2022mace, batatia2025design}, thus, to prove that TCE features are contained within MACE's feature space, it suffices to recover the second product using root-centered MACE features.

For each shell pair $(a,b)$, let $\Omega_{ab}$ be its finite set of admissible angle cosines. For some $\cos\theta\in\Omega_{ab}$, finite interpolation gives a polynomial $q_{ab}^{c}$ such that:

\begin{equation}\label{eq:outline-pair-projector}
    q_{ab}^{c}(\cos\theta)=\begin{cases}
        1 & d_a^2+d_b^2-2d_ad_b\cos\theta=d_c^2 \\
        0 & \text{else}
    \end{cases}
\end{equation}

where $a$, $b$, and $c$ denote neighbor shells and $d_a$, $d_b$, and $d_c$ are respectively the interatomic distance within those shells. Then, let $\mathbf{u}_\xi$ be the unit vector pointing from site $i$ to site $j_\xi$. We then have, by the law of cosines, that the neighbor-neighbor part of the adjacency tensor can be represented by the polynomial $q_{ab}^c$, given that $j_\xi$ and $j_\zeta$ are both connected to the root atom $i$:

\begin{equation}
    q_{a_\xi a_\zeta}^{c_{\xi\zeta}}(\mathbf{u}_\xi\cdot\mathbf{u}_\zeta) = A_{j_\xi j_\zeta}^{(c_{\xi\zeta})}
\end{equation}

so that the product

\begin{equation}\label{eq:outline-projector}
    Q_{\mathbf a}^{\mathbf c}(\mathbf u_1,\ldots,\mathbf u_\nu)= \prod_{\xi<\zeta} q_{a_\xi a_\zeta}^{c_{\xi\zeta}}(\mathbf u_\xi\cdot\mathbf u_\zeta)
\end{equation}

equals the neighbor-neighbor adjacency product in Eq.~\eqref{eq:outline-rooted} whenever the root-shell factors are nonzero. Furthermore, the function $Q_{\mathbf a}^{\mathbf c}$ is a finite-degree polynomial in pairwise dot products. It therefore has a finite expansion in scalar coupled spherical harmonics:

\begin{equation}\label{eq:outline-angular}
    Q_{\mathbf a}^{\mathbf c}=\sum_{\eta_\nu} \Lambda_{\eta_\nu}^{\mathbf a,\mathbf c}
    \mathcal K_{\eta_\nu}
\end{equation}

where $\eta_\nu$ is an angular-coupling multi-index, $K_{\eta_\nu}$ is an invariant angular kernel constructed as a product of spherical harmonics $Y_l^m$ and generalized Clebsch-Gordan coefficients $\mathcal{C}$:

\begin{equation}\label{eq:general-angular-kernel}
    \mathcal{K}_{\eta_\nu} \left( \widehat{\mathbf u}_1,\ldots,\widehat{\mathbf u}_\nu \right) = \sum_{\boldsymbol{lm}} \mathcal{C}_{\eta_\nu,\boldsymbol{lm}}^{00} \prod_{\xi=1}^{\nu} Y_{l_\xi}^{m_\xi}(\widehat{\mathbf u}_\xi)
\end{equation}

Here, consistent with MACE's notation, each $\eta_\nu$ labels a choice of angular degrees $(l_1, \cdots, l_\nu)$ yielding a total $L = 0$. Then, choose the primitive MACE channels to be

\begin{equation}\label{eq:outline-primitive}
    \mathcal A_{i,(a,\sigma)lm}^{(1)}=\sum_j A_{ij}^{(a)}X_{j\sigma} Y_l^m(\widehat{\mathbf r}_{ji})
\end{equation}

and let $\widehat B_{i,\eta_\nu;\mathbf p,00}^{(1)}$ denote the scalar
coupled product of these primitives, with
$p_\xi=(a_\xi,\sigma_\xi)$:

\begin{equation}\label{eq:slot-resolved-B}
    \widehat{B}_{i,\eta_\nu;\mathbf p,00}^{(1)}= \sum_{\boldsymbol{lm}} \mathcal{C}_{\eta_\nu,\boldsymbol{lm}}^{00} \prod_{\xi=1}^{\nu} \mathcal{A}_{i,p_\xi l_\xi m_\xi}^{(1)}
\end{equation}

Then, expanding the neighbor sums and using Eqs.~\eqref{eq:outline-projector}--\eqref{eq:outline-angular} gives:

\begin{equation}\label{eq:outline-local-identity}
    X_{i\alpha_1}\sum_{\eta_\nu}\Lambda_{\eta_\nu}^{\mathbf a,\mathbf c} \widehat B_{i,\eta_\nu;\mathbf p,00}^{(1)}=\Phi_{i,f}
\end{equation}

By the channel-realization condition, the finite combination in Eq.~\eqref{eq:outline-local-identity} can be implemented by the chosen MACE products. Receiving-species-dependent weights supply $X_{i\alpha_1}$, and a linear update sums the auxiliary channels, producing $h_{i,f00}^{(2)}=\Phi_{i,f}$, with a final summation layer yielding $N_f$. It follows immediately that a linear invariant readout recovers any
Hamiltonian in the corresponding finite TCE span:
\begin{equation}\label{eq:outline-readout}
    \begin{aligned}
        H &=\sum_{f\in\mathcal F}J_f\sum_i h_{i,f00}^{(2)}=\sum_{f\in\mathcal F}J_fN_f \\
        &\propto\sum_{\boldsymbol{\alpha}, \mathbf{a}, \mathbf{c}}\varepsilon_{\boldsymbol{\alpha}}^{\mathbf{a}\oplus\mathbf{c}} N_{\boldsymbol{\alpha}}^{\mathbf{a}\oplus\mathbf{c}}
    \end{aligned}
\end{equation}

Therefore:

\begin{equation}\label{eq:spanning-relation}
    \left\{N_{\boldsymbol{\alpha}}^{\ell'}\right\} \subseteq \operatorname{span} \left\{ \text{invariant MACE features} \right\}
\end{equation}

We note that the exact 1-to-1 equivalence described by the spanning relation above is not very important. Rather, what is important is that the two spaces are at least approximately equivalent in the information theory sense. A practical test for this is to see if, when computing a maximum entropy subset, that we can choose either a TCE or MACE feature vector and achieve the same subset. Then, this means that a TCE feature vector is an effective pre-screening surrogate for chemical complexity, allowing for a more efficient choice of a training set for a MACE potential, as done in this work.

\subsection{Full proof}\label{sec:full-proof}

We prove below that these generalized TCE features emerge as a subspace of MACE features \cite{batatia2022mace, batatia2025design}. More specifically, TCE features span an invariant subspace of MACE features with a particular choice of weights, and some assumptions about the message-passing architecture, which is tuneable by design. We can prove this by first constructing the non-symmetrized TCE feature $N_{\boldsymbol{\alpha}}^{\ell'}$:

\begin{equation}
    N_{\boldsymbol{\alpha}}^{\ell'}=\sum_{i_1,\ldots,i_m}\left[\prod_{1\leq r<s\leq m}A_{i_ri_s}^{(\ell'_{rs})}\right]\left[\prod_{r=1}^{m}X_{i_r\alpha_r}\right]
\end{equation}

Assume that the shell adjacency matrices are determined by the embedded
pair geometry on a fixed lattice, so that:

\begin{equation}
    A_{ij}^{(n)}=\chi_n(r_{ij}),   
\end{equation}

or, equivalently, that the shell labels are available to MACE as edge attributes. Then, for sufficiently large but finite angular cutoff and channel dimension, the feature $N_{\boldsymbol{\alpha}}^{\ell'}$ can be represented exactly by an invariant MACE feature of correlation order $\nu = m-1$. Without loss of generality, we can then pick an arbitrary lattice site, labeled the root site, $i_1$. Then, its neighbors can be written as:

\begin{equation}
    \begin{aligned}
        &i=i_1, \qquad j_\xi=i_{\xi+1}, \\
        &\xi=1,\ldots,\nu, \qquad \nu=m-1.
    \end{aligned}
\end{equation}

Here, $\xi$ iterates through the possible body orders, so that $j_\xi$ labels neighboring atoms that are directly connected to the root site.

Additionally, define the shell labels, which enumerate the neighbor orders:

\begin{equation}
    a_\xi:=\ell'_{1,\xi+1}
    \quad
    c_{\xi\zeta}:=\ell'_{\xi+1,\zeta+1}
    \quad
    1\leq \xi<\zeta\leq\nu
\end{equation}

and similarly the neighbor chemical labels, which enumerate the chemical types of $i_1$'s neighbors:

\begin{equation}
    \sigma_\xi:=\alpha_{\xi+1}.
\end{equation}

The local contribution rooted at $i_1$ is therefore
\begin{equation}\label{eq:tce-local-general}
    \begin{aligned}
        \Phi_{i,f} &= X_{i\alpha_1}\sum_{j_1,\ldots,j_\nu} \left[ \prod_{\xi=1}^{\nu} A_{ij_\xi}^{(a_\xi)} X_{j_\xi\sigma_\xi} \right] \left[ \prod_{1\leq\xi<\zeta\leq\nu} A_{j_\xi j_\zeta}^{(c_{\xi\zeta})} \right]
    \end{aligned}
\end{equation}

where $f=(\boldsymbol{a},\boldsymbol{c},\boldsymbol{\alpha})$ is a channel index and the corresponding global TCE feature is $N_f=\sum_i\Phi_{i,f}$. $\Phi_{i,f}$ then counts the number of clusters of type $f$ incident to site $i$. 

Note that the adjacency tensors are split into two different products, namely $A_{ij_\xi}^{(a_\xi)}$ and $A_{j_\xi j_\zeta}^{(c_{\xi\zeta})}$. The former represents the edges into site $i$, labeled root-neighbor edges, while the latter represents edges between neighbors of site $i$, labeled neighbor-neighbor edges. In the generalized formalism above, a cluster of type $f$ is fully defined by both, which is in contrast to the MACE formalism, whose many-body features are fully defined by the former, visualized in Fig.~\ref{fig:root-neighbor-v-neighbor-neighbor}.

We can engineer the weights of MACE, however, to include the contributions of the latter adjacency tensor. The standard MACE feature is \cite{batatia2022mace}:

\begin{equation}
    \mathcal{A}_{i,kl_1m_1}^{(1)} = \sum_{j\in\mathcal{N}(i)} R_{kl_1}^{(1)}(r_{ji})Y_{l_1}^{m_1}(\widehat{\mathbf{r}}_{ji}) W_{k\alpha_j}^{(1)}
\end{equation}

where $\mathcal{N}(i)$ is the set of neighbors of atom $i$, $R_{kl_1}^{(1)}$ is a radial basis function that depends on an implicit channel $k$, $r_{ji}$ is the distance between atoms $i$ and $j$, $Y_{l_1}^{m_1}$ denotes spherical harmonics, $\widehat{\mathbf{r}}_{ji}$ is the unit vector between atoms $i$ and $j$, $\alpha_j$ is the chemical type of atom $j$, and $W_{k\alpha_j}^{(1)}$ is a species-resolved weight. We note that MACE uses a standard $A$ to denote this feature, and additionally denotes the chemical species by their atomic number $z_j$. For this work, to simplify the connection to TCE features, we will use the alternative notation above, namely $\mathcal{A}$ for MACE features and Greek letters for chemical species.

We can design the radial basis functions as well as the species-resolved weight to match TCE descriptors in Eq.~\eqref{eq:tce-local-general}, namely by choosing the channel $k = (n,\sigma)$ to correspond to neighbor order $n$ and chemical type $\sigma$, and choosing a radial basis function:

\begin{equation}
    \begin{aligned}
        \chi_n(d_{n'}) &= \delta_{nn'} \\
        R_{(n,\alpha)}^{(1)}(r) &= \chi_n(r) \\
        W_{(n,\alpha)\beta}^{(1)} &= \delta_{\alpha \beta}
    \end{aligned}
\end{equation}

where $d_{n}$ denotes the neighbor distance for shell $n$ so that $\chi_n$ is a shell indicator. In general, this should be continuous over $r$, i.e. a shell indicator if $d_n$ is exactly a neighbor shell, and an interpolant otherwise. These choices yield  shell- and species-resolved features that satisfy:
\begin{equation}\label{eq:mace-A-shell-species}
    \mathcal{A}_{i,(n,\sigma)lm}^{(1)}= \sum_j A_{ij}^{(n)} X_{j\sigma} Y_l^m(\widehat{\mathbf r}_{ji}).
\end{equation}

where $A_{ij}^{(n)}$ is the adjacency tensor defined in Eq.~\eqref{eq:topology-tensors-old}. For the target TCE feature, define the primitive channels:

\begin{equation}
    p_\xi:=(a_\xi,\sigma_\xi) \qquad \xi=1,\ldots,\nu
\end{equation}

where $a_\xi$ denotes a neighbor order, i.e. the edge type connecting atoms $i$ and $j_\xi$, and $\sigma_\xi$ is the chemical type of atom $j_\xi$. Eq.~\eqref{eq:mace-A-shell-species} then gives:

\begin{equation}\label{eq:mace-A-primitive}
    \mathcal{A}_{i,p_\xi l_\xi m_\xi}^{(1)}=\sum_{j_\xi} A_{ij_\xi}^{(a_\xi)} X_{j_\xi\sigma_\xi} Y_{l_\xi}^{m_\xi} \bigl(\widehat{\mathbf r}_{j_\xi i}\bigr).
\end{equation}

Introduce the slot-resolved coupled product
\begin{equation}
\label{eq:slot-resolved-B-appendix}
\widehat{B}_{i,\eta_\nu;\mathbf p,00}^{(1)}= \sum_{\boldsymbol{lm}} \mathcal{C}_{\eta_\nu,\boldsymbol{lm}}^{00} \prod_{\xi=1}^{\nu} \mathcal{A}_{i,p_\xi l_\xi m_\xi}^{(1)},
\end{equation}

where:

\begin{equation}
    \mathbf p=(p_1,\ldots,p_\nu) \qquad \boldsymbol{lm} = (l_1m_1,\ldots,l_\nu m_\nu)
\end{equation}

and $\eta_\nu$ labels a generalized Clebsch-Gordan coupling path to total angular momentum $L=M=0$, and $\mathcal{C}$ denotes the generalized Clebsch-Gordan coefficients. Here, the coupling and the coefficients themselves retain the same meaning as within the standard MACE formalism. Substituting Eq.~\eqref{eq:mace-A-primitive} into
Eq.~\eqref{eq:slot-resolved-B-appendix} yields:

\begin{equation}\label{eq:B-expanded-general}
\begin{aligned}
    \widehat{B}_{i,\eta_\nu;\mathbf p,00}^{(1)} = \sum_{j_1,\ldots,j_\nu} \left[ \prod_{\xi=1}^{\nu} A_{ij_\xi}^{(a_\xi)} X_{j_\xi\sigma_\xi} \right] \mathcal{K}_{\eta_\nu} \left( \widehat{\mathbf r}_{j_1i}, \ldots, \widehat{\mathbf r}_{j_\nu i} \right)
\end{aligned}
\end{equation}

where the invariant angular kernel is:

\begin{equation}
    \mathcal{K}_{\eta_\nu} \left( \widehat{\mathbf u}_1,\ldots,\widehat{\mathbf u}_\nu \right) = \sum_{\boldsymbol{lm}} \mathcal{C}_{\eta_\nu,\boldsymbol{lm}}^{00} \prod_{\xi=1}^{\nu} Y_{l_\xi}^{m_\xi}(\widehat{\mathbf u}_\xi)
\end{equation}

Thus, the product of the $\nu$ primitive $\mathcal{A}$ features already produces the target factors $ \prod_{\xi=1}^{\nu} A_{ij_\xi}^{(a_\xi)} X_{j_\xi\sigma_\xi}$ namely the $\nu$ root-neighbor adjacency factors and the $\nu$ neighbor occupation factors.

What remains is to construct the $\binom{\nu}{2}$ adjacency factors among the neighbors, i.e. the neighbor-neighbor edges visualized in Fig.~\ref{fig:root-neighbor-v-neighbor-neighbor}, which would complete the proof. For fixed root-shell labels $a$ and $b$, define the finite set of the cosine of admissible lattice angles:

\begin{equation}
    \Omega_{ab} = \left\{ \widehat{\mathbf r}_{ji}\cdot \widehat{\mathbf r}_{ki} \;|\; A_{ij}^{(a)}A_{ik}^{(b)}=1 \right\}
\end{equation}

For every closing shell $c$ (closing in the sense that the $(a,b,c)$ edge forms a closed clique), choose a finite interpolation polynomial $q_{ab}^{c}$ satisfying:
\begin{equation} \label{eq:q-projector-general}
    q_{ab}^{c}(\cos\theta) = \begin{cases}
        1 & d_a^2+d_b^2-2d_ad_b\cos\theta=d_c^2 \\
        0 & \text{else}
    \end{cases}
\end{equation}

for all $\cos\theta\in\Omega_{ab}$, where $d_a$ is the distance in shell $a$ and $\theta$ is the angle opposite to $\{a, b\}$ side in the triangle defined by the nodes $\{a,b,c\}$. Then, define the complete closure-graph projector:

\begin{equation}\label{eq:closure-projector-general}
\begin{aligned}
    Q_{\mathbf a}^{\mathbf c} \left(\widehat{\mathbf u}_1,\ldots,\widehat{\mathbf u}_\nu\right)= \prod_{1\leq\xi<\zeta\leq\nu} q_{a_\xi a_\zeta}^{c_{\xi\zeta}} \left( \widehat{\mathbf u}_\xi\cdot \widehat{\mathbf u}_\zeta \right)
\end{aligned}
\end{equation}

Conditional on the root-shell factors,
Eq.~\eqref{eq:q-projector-general} and the law of cosines imply:

\begin{equation}\label{eq:closure-projector-identity}
\begin{aligned}
    \left[\prod_{\xi=1}^{\nu}A_{ij_\xi}^{(a_\xi)}\right]&Q_{\mathbf a}^{\mathbf c}\left(\widehat{\mathbf r}_{j_1i}, \ldots, \widehat{\mathbf r}_{j_\nu i} \right)\\
    &=\left[\prod_{\xi=1}^{\nu}A_{ij_\xi}^{(a_\xi)}\right]\left[\prod_{\xi<\zeta} A_{j_\xi j_\zeta}^{(c_{\xi\zeta})} \right]
\end{aligned}
\end{equation}

Note that, when $a_\xi=a_\zeta$, repeated-neighbor contributions $j_\xi=j_\zeta$ may be removed by including $\mu=1$ in the
interpolation set and imposing $q_{aa}^{c}(1)=0$ for every positive closing shell $c$.

The function $Q_{\mathbf a}^{\mathbf c}$ is a finite polynomial in the pairwise dot products
$\widehat{\mathbf u}_\xi\cdot\widehat{\mathbf u}_\zeta$. It is therefore a finite-degree, rotationally invariant polynomial on
$(S^2)^\nu$. Hence, for sufficiently large but finite angular cutoff, it has an expansion in coupled spherical harmonics with total $L=M=0$:

\begin{equation}\label{eq:Q-coupled-expansion}
    Q_{\mathbf a}^{\mathbf c}\left( \widehat{\mathbf u}_1,\ldots,\widehat{\mathbf u}_\nu \right) = \sum_{\eta_\nu} \Lambda_{\eta_\nu}^{\mathbf a,\mathbf c} \mathcal{K}_{\eta_\nu} \left( \widehat{\mathbf u}_1,\ldots,\widehat{\mathbf u}_\nu \right)
\end{equation}

Combining Eqs.~\eqref{eq:B-expanded-general},
\eqref{eq:closure-projector-identity}, and
\eqref{eq:Q-coupled-expansion} gives:

\begin{equation}\label{eq:B-produces-all-neighbor-factors}
\begin{aligned}
    \sum_{\eta_\nu} &\Lambda_{\eta_\nu}^{\mathbf a,\mathbf c} \widehat{B}_{i,\eta_\nu;\mathbf p,00}^{(1)}\\
    &= \sum_{j_1,\ldots,j_\nu} \left[ \prod_{\xi=1}^{\nu} A_{ij_\xi}^{(a_\xi)} X_{j_\xi\sigma_\xi} \right] \left[ \prod_{\xi<\zeta} A_{j_\xi j_\zeta}^{(c_{\xi\zeta})} \right]
\end{aligned}
\end{equation}

The slot-resolved object $\widehat{B}_{i,\eta_\nu;\mathbf p,00}^{(1)}$ defined in Eq.~\eqref{eq:slot-resolved-B-appendix} is not itself written in the factorized channel notation of the native MACE $B$-features:

\begin{equation}
    B_{i,\eta_\nu k00}^{(1)} = \sum_{\boldsymbol{lm}} C_{\eta_\nu,\boldsymbol{lm}}^{00}\prod_{\xi=1}^\nu \sum_{\tilde{k}} w_{k\tilde{k}l_\xi}^{(1)} A_{i,\tilde{k}l_\xi m_\xi}^{(1)}
\end{equation}

We could, however, choose the weight $w_{k\tilde{k}l_\xi}^{(1)} = \delta_{k\tilde{k}}$ so that the MACE $B$-features exactly match the slot-resolved coupled product. More generally, though, the required permutation-symmetrized channel tensor lies in the finite linear span of the native features $B_{i,\eta_\nu k00}^{(1)}$. Notably, the neighbor indices are dummy summation variables, so the channel tensor may be symmetrized, and every finite symmetric order-$\nu$ tensor is a finite linear combination of rank-one symmetric powers \cite{comon2008symmetric}. Consequently, there exist finitely many native MACE channels $k_q$ and coefficients $\Theta_{q,\eta_\nu}^{f}$ such that:

\begin{equation}
\label{eq:native-MACE-realization}
    \sum_{\eta_\nu} \Lambda_{\eta_\nu}^{\mathbf a,\mathbf c} \widehat{B}_{i,\eta_\nu;\mathbf p,00}^{(1)} = \sum_q\sum_{\eta_\nu} \Theta_{q,\eta_\nu}^{f} B_{i,\eta_\nu k_q00}^{(1)}
\end{equation}

Choose the receiving-species-dependent MACE message weights as
\begin{equation}\label{eq:central-species-weights}
    W_{\alpha_i k_q0,\eta_\nu}^{(1)} = \delta_{\alpha_i\alpha_1} \Theta_{q,\eta_\nu}^{f} = X_{i\alpha_1} \Theta_{q,\eta_\nu}^{f}
\end{equation}

Set all message weights at correlation orders different from $\nu=m-1$ to zero. The resulting messages are
\begin{equation} \label{eq:general-message}
    m_{i,k_q00}^{(1)} = X_{i\alpha_1} \sum_{\eta_\nu} \Theta_{q,\eta_\nu}^{f} B_{i,\eta_\nu k_q00}^{(1)}
\end{equation}

Choose the subsequent MACE update to be linear, set the residual
weights to zero, and sum the auxiliary channels:
\begin{equation}\label{eq:general-linear-update}
h_{i,f00}^{(2)} = \sum_q m_{i,k_q00}^{(1)}
\end{equation}

Using Eqs.~\eqref{eq:B-produces-all-neighbor-factors} and \eqref{eq:native-MACE-realization}, this gives:

\begin{equation}\label{eq:node-feature-equals-tce}
\begin{aligned}
    h_{i,f00}^{(2)}&=X_{i\alpha_1}\sum_{j_1,\ldots,j_\nu}\left[\prod_{\xi=1}^{\nu}A_{ij_\xi}^{(a_\xi)} X_{j_\xi\sigma_\xi} \right] \left[ \prod_{\xi<\zeta} A_{j_\xi j_\zeta}^{(c_{\xi\zeta})} \right]\\
    &= \Phi_{i,f}
\end{aligned}
\end{equation}

Therefore:
\begin{equation}\label{eq:global-feature-equality}
    \sum_i h_{i,f00}^{(2)} = \sum_i\Phi_{i,f} = N_f
\end{equation}

A linear invariant readout with coefficient $J_f$ then gives:

\begin{equation} \label{eq:hamiltonian-readout}
    H = \sum_f J_f\sum_i h_{i,f00}^{(2)} = \sum_fJ_fN_f
\end{equation}

Finally, the number of adjacency factors represented by the correlation-order-$\nu$ MACE feature is:

\begin{equation}\label{eq:adjacency-count-general}
    \nu + \binom{\nu}{2} = \binom{\nu+1}{2} = \binom{m}{2}
\end{equation}

where $\nu$ is the number of root-neighbor edges and $\binom{\nu}{2}$ is the number of neighbor-neighbor edges. Likewise, the number of occupation factors $\mathbf{X}$ is:

\begin{equation}\label{eq:occupation-count-general}
     \nu+1 = m
\end{equation}

where $\nu$ is the number of $\mathbf{X}$ factors corresponding to neighbors and $1$ is the number of $\mathbf{X}$ factors corresponding to the occupation of the root atom. Hence every $m$-body TCE feature is exactly representable by an invariant MACE feature with correlation order $\nu=m-1$, therefore completing the proof, so:

\begin{equation}
    \left\{N_{\boldsymbol{\alpha}}^{\ell'}\right\} \subseteq \operatorname{span} \left\{ \text{invariant MACE features} \right\}
\end{equation}

Note that these features are of the non-symmetrized cluster counts $\ell'$, rather than the symmetrized counts indexed by the equivalence class $[\ell]$. For a symmetric TCE environment $[\ell]$, the corresponding topology
tensor is:

\begin{equation}
    A_{i_1\cdots i_m}^{[\ell]}= \bigvee_{\ell'\in[\ell]} \prod_{1\leq r<s\leq m} A_{i_ri_s}^{(\ell'_{rs})}
\end{equation}

Since distinct complete shell assignments are mutually exclusive on a fixed site tuple, the Boolean union is proportional to the ordinary sum:

\begin{equation}
    A_{i_1\cdots i_m}^{[\ell]}\propto \sum_{\ell'\in[\ell]} \prod_{1\leq r<s\leq m} A_{i_ri_s}^{(\ell'_{rs})}
\end{equation}

where the coefficient of proportionality can depend on the stabilizer of $\ell$ under the action of $S_m$. Thus the feature $N_{\boldsymbol{\alpha}}^{[\ell]}$ is obtained by a final linear sum over the MACE channels constructed for the distinct representatives $\ell'\in[\ell]$.

In summary, the proof above shows that we can, for any permissible value of $\nu$, engineer the spherical harmonics within the MACE features to index neighbor-neighbor edges. For two-body terms, where there are no neighbor-neighbor edges, the construction is trivial, i.e. fix the spherical harmonics to be constant. For higher-order terms, the construction relies on the fact that $\mathcal{K}_l$ is $\text{SO}(3)$-invariant over $(S^2)^\nu$, ensuring the existence of abstract weights that match the TCE Hamiltonian. For convenience, we have included explicit constructions for two and three-body terms in Sec.~\ref{sec:two-three-bodies}.

The construction above does not establish an equivalence between the full TCE and MACE feature spaces. Instead, it shows that every TCE feature belongs to a particular shell-resolved subspace of the MACE feature space. In this subspace, the radial channels act as pair-shell projectors, while the angular couplings retain only the information required to distinguish complete species-decorated shell topologies.

To formalize this statement, let $\mathcal{U}_m$ denote the space of possible m-body environments. The completeness results of Dusson et al. show that the ACE basis systematically spans the relevant space of permutation- and isometry-invariant functions on $\mathcal{E}_m$ \cite{dusson2022atomic}, subject to the usual basis-completeness and regularity assumptions. MACE constructs its higher-order messages from the same type of ACE product basis \cite{drautz2019atomic}.

Within TCE, two environments $u, u'\in\mathcal{U}_m$ are equivalent when they have the same species-decorated pair-shell topology. Namely, if $\Gamma_m(u)$ is defined to be $K_m$ with labels for species and edge types, where $u = (\boldsymbol{\alpha}, \{\ell_{rs}\}_{r<s})$:

\begin{equation}
    \Gamma_m(u) = (K_m, \boldsymbol{\alpha}, \{\ell_{rs}\}_{r<s})
\end{equation}

then two environments are equivalent within TCE if the two labeled subgraphs are isomorphic:

\begin{equation}
    u\sim u' \iff \Gamma_m(u) \cong \Gamma_m(u')
\end{equation}

where we define two labeled graphs $G = (V, E)$ and $G' = (V', E')$ as isomorphic if there exists permutations $\pi_1$ and $\pi_2$ such that $V' = \pi_1 V$ and $E' = \pi_2 E$. Then, the quotient $\mathcal{Q}_m = \mathcal{U}_m/\sim$ is the set of distinct $m$-body environments that TCE can distinguish. We can then define a basis of indicator functions for one class $[q]\in\mathcal{Q}_m$:

\begin{equation}
    \tau_q(u) = \begin{cases}
        1 & u \sim q \\
        0 & \text{else}
    \end{cases}
\end{equation}

These indicators then satisfy the orthogonality relation $\tau_q(u)\tau_{q'}(u) = \delta_{qq'}\tau_q(u)$. Any function on $\mathcal{Q}_m$ can therefore be written as a linear combination of $\tau_q$, implying that TCE forms a complete basis over $\mathcal{Q}_m$.

\subsection{Two/three-body examples}\label{sec:two-three-bodies}

Below we show specific cases of the described equivalence, namely the case for two-body and three-body features.

\subsubsection{Two-body Hamiltonian}

Starting from Eq.~\eqref{eq:mace-A-shell-species}, fixing $l = m = 0$ gives us:

\begin{equation}
    \mathcal{A}_{i,(n,\alpha)00}^{(1)} = \frac{1}{\sqrt{4\pi}}\sum_j A_{ij}^{(n)} X_{j\alpha}
\end{equation}

This is already quite close to the TCE descriptors. Namely, $\mathcal{A}_{i,(n,\alpha)00}^{(1)}$ is proportional to the number of $n$'th nearest neighbors of atom $i$ that are of type $\alpha$. By fixing the weights, we can additionally attach the type of atom $i$ to the message. First, we can set $\nu = 1$, fixing the body order $\nu + 1 = 2$, yielding the feature tensor:

\begin{equation}
    B_{i,\eta_1k00}^{(1)} = \sum_{\tilde{k}} w_{k\tilde{k}0}^{(1)}\mathcal{A}_{i\tilde{k}00}^{(1)}
\end{equation}

If here we introduce the channel $k = (n,\beta,\alpha)$, corresponding to neighbor order $n$ and chemical types $\alpha$ and $\beta$, we can select the weights to be orthogonal:

\begin{equation}
    w_{(n,\beta,\alpha),(n',\alpha'),0}^{(1)} = \delta_{nn'}\delta_{\alpha\alpha'}
\end{equation}

Such that:

\begin{equation}
    B_{i,\eta_1,(n,\beta,\alpha),00}^{(1)} = \mathcal{A}_{i,(n,\alpha)00}^{(1)}
\end{equation}

which is proportional to the number of $\alpha$ atoms that are $n$'th nearest neighbors of atom $i$. The message into atom $i$ is then:

\begin{equation}
    \begin{aligned}
        m_{i,k00}^{(1)} &= W_{\alpha_ik0,\eta_1}^{(1)}B_{i,\eta_1k00}^{(1)}\\
    \implies &m_{i,(n,\beta,\alpha)00}^{(1)} = W_{\alpha_i(n,\beta,\alpha)0,\eta_1}^{(1)}B_{i,\eta_1(n,\beta,\alpha)00}^{(1)} 
    \end{aligned}
\end{equation}

where $\alpha_i$ is the chemical type of atom $i$. Choose the weight $W_{\gamma(n,\beta,\alpha)0,\eta_1} = \delta_{\gamma\beta}$. Then, $W_{\alpha_i,(n,\beta,\alpha),0,\eta_1} = \delta_{\alpha_i\beta} = X_{i\beta}$, so:

\begin{equation}
    m_{i,k00}^{(1)} = X_{i\beta} \mathcal{A}_{i,(n,\alpha)00}^{(1)} = \frac{1}{\sqrt{4\pi}}\sum_j A_{ij}^{(n)}X_{j\alpha}X_{i\beta}
\end{equation}

The atomic energy is then ordinarily computed via its own MLP, namely:

\begin{equation}
    E_i = f_i(m_{i,k00}^{(1)}) = \sum_k W_{ik}m_{i,k00}^{(1)}
\end{equation}

where I've implicitly set that atom's MLP to a single linear readout. I will make the assumption here that every atom shares the same network, namely there's one weight $W_{ik} = W_k$. So:

\begin{equation}
    \begin{aligned}
        H &= \sum_i E_i = \sum_{ik} W_{ik}m_{i,k00}^{(1)} = \sum_k W_k\sum_i m_{i,k00}^{(1)}\\
        &= \frac{1}{\sqrt{4\pi}}\sum_{(n,\beta,\alpha)}W_{(n,\beta,\alpha)}^{(1)}\sum_{ij}A_{ij}^{(n)}X_{j\alpha}X_{i\beta} \\
        &=\frac{1}{\sqrt{4\pi}}\sum_{(n,\alpha,\beta)} W_{(n,\alpha,\beta)}^{(1)} \sum_{ij}A_{ij}^{(n)}X_{i\alpha} X_{j\beta}
    \end{aligned}
\end{equation}

I.e. the Hamiltonian for this MACE architecture is equivalent to the TCE hamiltonian, with linear readout weights proportional to the interaction energies $\varepsilon_{\alpha\beta}^{(n)}$.

\subsubsection{Three-body Hamiltonian}

Take $\nu=2$ for three body terms. Then, the MACE feature tensor is:

\begin{equation}
    \begin{aligned}
        B_{i,\eta_2kLM}^{(1)} &= \sum_{m_1,m_2,l_1,l_2} \mathcal{C}_{\eta_2, l_1m_1, l_2m_2}^{LM}\\
    &\left[ \sum_{\tilde{k}} w_{k\tilde{k}l_1}^{(1)}\mathcal{A}_{i,\tilde{k}l_1m_1}^{(1)}\right] \left[ \sum_{\tilde{k}} w_{k\tilde{k}l_2}^{(1)}\mathcal{A}_{i,\tilde{k}l_2m_2}^{(1)}\right]
    \end{aligned}
\end{equation}

For a scalar feature (scalar in the tensor sense), take $L = M = 0$ so that $l = l_1 = l_2$. Then:

\begin{equation}
    \begin{aligned}
        B_{i,\eta_2(l)k00}^{(1)} = &\sum_{m_1,m_2} \mathcal{C}_{\eta_2(l), lm_1, lm_2}^{00}\\
        &\left[ \sum_{\tilde{k}} w_{k\tilde{k}l}^{(1)}\mathcal{A}_{i,\tilde{k}lm_1}^{(1)}\right] \left[ \sum_{\tilde{k}} w_{k\tilde{k}l}^{(1)}\mathcal{A}_{i,\tilde{k}lm_2}^{(1)}\right]
    \end{aligned}
\end{equation}

where $\eta_2$ depends on $l$. We choose a channel $\tilde{k} = (n,\sigma)$ once again to match the neighbor shell $n$ and the chemical type $\sigma$. As shown in the two-body features, this lets us construct the $\mathcal{A}$ features:

\begin{equation}
    \mathcal{A}_{i,(n,\sigma)lm}^{(1)} = \sum_j A_{ij}^{(n)} X_{j\sigma} Y_l^m (\hat{\mathbf{r}}_{ji})
\end{equation}

Then, if $p = (a,\alpha)$ and $q = (b,\gamma)$:

\begin{equation}
    \begin{aligned}
        \mathcal{A}_{i,plm}^{(1)} &= \sum_{j_1} A_{ij_1}^{(a)} X_{j_1\alpha}Y_l^m(\hat{\mathbf{r}}_{j_1i}) \\
        \mathcal{A}_{i,qlm}^{(1)} &= \sum_{j_2} A_{ij_2}^{(b)} X_{j_2\gamma} Y_l^m(\hat{\mathbf{r}}_{j_2i})
    \end{aligned}
\end{equation}

Choose the weights $w_{k\tilde{k}l}^{(1)}$, i.e. let $k = \pm 1$:

\begin{equation}
    w_{k_\pm,\tilde{k},l}^{(1)} = \delta_{\tilde{k}, p} \pm \delta_{\tilde{k}, q}
\end{equation}

Then:

\begin{equation}
    \sum_{\tilde{k}} w_{k_\pm,\tilde{k},l}^{(1)}\mathcal{A}_{i,\tilde{k}lm}^{(1)} = \mathcal{A}_{i,plm}^{(1)}\pm \mathcal{A}_{i,qlm}^{(1)}
\end{equation}

Plugging into the $B$ feature tensor:

\begin{equation}
    \begin{aligned}
        B_{i\eta_2,k_\pm 00}^{(1)} = &\sum_{m_1,m_2} \mathcal{C}_{\eta_2, lm_1, lm_2}^{LM}\\
        &\left[ \mathcal{A}_{i,plm_1}^{(1)}\pm \mathcal{A}_{i,qlm_1}^{(1)}\right] \left[ \mathcal{A}_{i,plm_2}^{(1)}\pm \mathcal{A}_{i,qlm_2}^{(1)}\right]
    \end{aligned}
\end{equation}

If we are linearly fitting to the tensors above, we can equivalently linearly fit to any linear combination of the tensors. So, we can take the difference:

\begin{equation}
    \begin{aligned}
        D_{i,l}^{pq} &= \frac{B_{i\eta_2,k_+ 00}^{(1)} - B_{i\eta_2,k_- 00}^{(1)}}{4}\\
        &=\frac{1}{2}\sum_{m_1,m_2}\mathcal{C}_{\eta_2, lm_1, lm_2}^{00}\\
        &\left( \mathcal{A}_{i,qlm_1}^{(1)}\mathcal{A}_{i,plm_2}^{(1)}+ \mathcal{A}_{i,plm_1}^{(1)} \mathcal{A}_{i,qlm_2}^{(1)}\right)
    \end{aligned}
\end{equation}

The generalized Clebsch-Gordan coefficients satisfy the exchange symmetry:

\begin{equation}
    \mathcal{C}_{\eta_2,lm_2,lm_1}^{LM} = (-1)^{2l-L}\mathcal{C}_{\eta_2,lm_1lm_2}^{LM}
\end{equation}

So:

\begin{equation}
    \begin{aligned}
        D_{i,l}^{pq} &= \frac{B_{i\eta_2,k_+ 00}^{(1)} - B_{i\eta_2,k_- 00}^{(1)}}{4} \\
        &=\sum_{m_1,m_2}\mathcal{C}_{\eta_2, lm_1, lm_2}^{00}\mathcal{A}_{i,plm_1}^{(1)}\mathcal{A}_{i,qlm_2}^{(1)}
    \end{aligned}
\end{equation}

Then, substituting in the two $\mathcal{A}$ features:

\begin{equation}
    \begin{aligned}
        D_{i,l}^{pq} &= \frac{B_{i\eta_2,k_+ 00}^{(1)} - B_{i\eta_2,k_- 00}^{(1)}}{4}\\
        &=\sum_{j_1, j_2} A_{ij_1}^{(a)} A_{ij_2}^{(b)}X_{j_1\alpha} X_{j_2\gamma} \\
        &\times\sum_{m_1,m_2}\mathcal{C}_{\eta_2, lm_1, lm_2}^{00}Y_l^{m_1}(\hat{\mathbf{r}}_{j_1i})Y_l^{m_2}(\hat{\mathbf{r}}_{j_2i})
    \end{aligned}
\end{equation}

Defining the scalar kernel:

\begin{equation}
    \mathcal{K}_l(\hat{\mathbf{r}}, \hat{\mathbf{r}}') = \sum_{m_1,m_2}\mathcal{C}_{\eta_2, lm_1, lm_2}^{00}Y_l^{m_1}(\hat{\mathbf{r}})Y_l^{m_2}(\hat{\mathbf{r}}')
\end{equation}

which, by the addition theorem, simplifies to:

\begin{equation}
    \mathcal{K}_l(\hat{\mathbf{r}}, \hat{\mathbf{r}}') = \kappa_l P_l(\hat{\mathbf{r}}\cdot\hat{\mathbf{r}}')
\end{equation}

for some constant $\kappa_l$ and Legendre polynomials $P_l$. We hope here that we can reconstruct the remaining adjacency tensor $A_{j_1j_2}^{(c)}$ from the Legendre polynomials. We can then choose coefficients $\Lambda_l^{abc}$ such that, whenever $A_{ij_1}^{(a)}A_{ij_2}^{(b)} = 1$, we have:

\begin{equation}
    \sum_l \Lambda_l^{abc} \mathcal{K}_l(\hat{\mathbf{r}}_{j_1i}, \hat{\mathbf{r}}_{j_2i}) = A_{j_1j_2}^{(c)}
\end{equation}

We can construct these coefficients by lettitng $\mu_s = \hat{\mathbf{r}}_{j_1i}\cdot\hat{\mathbf{r}}_{j_2i}$ denote the possible angles indexed by $s$, which is finite in a fixed lattice, and define the matrix elements $G_{sl}^{ab} = \mathcal{K}_l(\mu)$, as well as the indicator tensor for the third leg:

\begin{equation}
    y_s^{abc} = \begin{cases}
        1 & d_a^2 + d_b^2 - 2d_ad_b \mu_s = d_c^2 \\
        0 & \text{else}
    \end{cases}
\end{equation}

\begin{equation}
    \sum_l G_{sl}^{ab}\Lambda_l^{abc} = y_s^{abc}
\end{equation}

which is, fortunately, invertible, since $l$ is strictly finite here, i.e. $l=0,\cdots,S_{ab} - 1$ where $S_{ab}$ is the number of distinct values of $s$. This is alternatively a Green's function. Then:

\begin{equation}
    \Lambda_l^{abc} = (G^{ab})^{-1} y^{abc}
\end{equation}

The message into atom $i$ is then, once again:

\begin{equation}
    m_{i,kLM}^{(1)} = \sum_{\nu,\eta_\nu} W_{\alpha_i kL,\eta_\nu}^{(1)} B_{i,\eta_\nu kLM}^{(1)}
\end{equation}

Choose the weights for the two channels we defined earlier, i.e. $k=\pm 1$, $L = 0$ for even parity, and $\nu = 2$:

\begin{equation}
    W_{\alpha_ikL,\eta_2} = \delta_{\alpha_i\beta}\Lambda_l^{abc} = X_{i\beta}\Lambda_l^{abc}
\end{equation}

Then, the message simplifies to:

\begin{equation}
    m_{i,k_\pm,00} = X_{i\beta}\sum_l \Lambda_l^{abc} B_{i\eta_2(l)k_\pm 00}
\end{equation}

Here, $\eta_\nu$ implicitly depends on $l$, so we swap the sum to $l$. The updated feature from passing the message into a second layer is, with zero residual connections:

\begin{equation}
    h_{i,kLM}^{(2)} = \sum_{\tilde{k}}W_{kL,\tilde{k}}^{(1)}m_{i,\tilde{k}LM}^{(1)}
\end{equation}

With our chosen architecture, where $k = \pm 1$ and $L = M = 0$:

\begin{equation}
    h_{i,f00}^{(2)} = \sum_{k}W_{f0,k}^{(1)}m_{i,k00}^{(1)}
\end{equation}

Choose the weights $W_{f0,k_\pm} = \pm\frac{1}{4}$. Then:

\begin{equation}
    \begin{aligned}
        h_{i,f00}^{(2)} &= \frac{1}{4}m_{i,k_+00}^{(1)} -\frac{1}{4}m_{i,k_-00}^{(1)}\\
        &= X_{i\beta}\sum_l \Lambda_l^{abc} D_{i,l}^{pq}\\
        &= X_{i\beta}\sum_{j_1,j_2}A_{ij_1}^{(a)}A_{ij_2}^{(b)}X_{j_1\alpha}X_{j_2\gamma}\\
        &\times \sum_l \Lambda_l^{abc}\mathcal{K}_l(\hat{\mathbf{r}}_{j_1i},\hat{\mathbf{r}}_{j_2i})
    \end{aligned}
\end{equation}

By construction, the latter is exactly $A_{j_1j_2}^{(c)}$. Therefore:

\begin{equation}
    \begin{aligned}
        h_{i,f00}^{(2)} &= \sum_{j_1,j_2}A_{ij_1}^{(a)}A_{ij_2}^{(b)}A_{j_1j_2}^{(c)}X_{j_1\alpha}X_{j_2\gamma} X_{i\beta}\\
    &= \sum_{j,k}A_{ij}^{(a)}A_{ik}^{(b)}A_{jk}^{(c)}X_{j\alpha}X_{k\gamma} X_{i\beta}
    \end{aligned}
\end{equation}

$f$ represents the flattened index $f = (a,b,c;\alpha,\beta,\gamma)$. TCE also implicitly uses this indexing flattening when fitting. Then, similar to before, if we have a constant aggregation layer, i.e. a summation layer, into a linear readout layer, the Hamiltonian is:

\begin{equation}
    \begin{aligned}
        H &= \sum_{if}W_{if}^{(2)}h_{i,f00}^{(2)} = \sum_f W_f^{(2)}\sum_i h_{i,f00}^{(2)}\\
        &= \sum_{abc\alpha\beta\gamma} W_{abc\alpha\beta\gamma}^{(2)} \sum_{ijk}A_{ij}^{(a)}A_{ik}^{(b)}A_{jk}^{(c)}X_{j\alpha}X_{k\gamma} X_{i\beta}
    \end{aligned}
\end{equation}

which, if the weights are symmetric, is the exact TCE Hamiltonian. Namely, if:

\begin{equation}
    W_{abc\alpha\beta\gamma}^{(2)} = W_{(abc)\alpha\beta\gamma}^{(2)} = W_{abc(\alpha\beta\gamma)}^{(2)}
\end{equation}

then $W^{(2)}$ corresponds exactly to the ECIs $\varepsilon_{\alpha\beta\gamma}^{[\ell]}$, where $\ell = (a, b, c)$ is the three-body label. This symmetrization could be a separate linear layer, i.e. symmetrization is a linear operation, so this would correspond to another architecture choice.

\newpage

\subsection{$\text{W}\text{S}_2$-like clusters}

In Tab.~\ref{tab:mx2-many-body-clusters}, we describe the TCE features used to decorate the $\text{W}\text{S}_2$-like clusters. Namely, we include up to $4$-body interactions. Each body order has a finite set of topological labels $\ell$, which index edge types, and each feature is computed for the induced equivalence class $\ell\in S_m\cdot\ell = [\ell]$. We index edge types simply by nearest neighbor distances, accounting for periodic boundary conditions. For example, $m = 2$ and $\ell = 1$ correspond to the pairwise, nearest neighbor interaction. Strictly speaking, $\ell$ should be a tuple rather than an integer, but we can express singleton $\ell$'s as integers without loss of generality.

\begin{table*}[t]
    \centering
    \begin{tabularx}{\textwidth}{@{} l l l X @{}}
    \toprule
    number of bodies $m$ & topology labels $\ell$ & Species & Geometry \\
    \midrule
    $m = 2$ & $1$ & $\mathrm{WS}$ & First nearest neighbor, $\SI{2.5}{\AA} \pm 5\%$ \\
            & $2$ & $\mathrm{W}_2/\mathrm{S}_2$ & Second nearest neighbor, $\SI{3.25}{\AA} \pm 5\%$ \\
            & $3$ & $\mathrm{WS}$ & Third nearest neighbor, $\SI{4.0}{\AA} \pm 5\%$ \\
            & $4$ & $\mathrm{S}_2$ & Fourth nearest neighbor, $\SI{4.5}{\AA} \pm 5\%$ \\
            & $5$ & $\mathrm{WS}$ & Fifth nearest neighbor, $\SI{5.0}{\AA} \pm 5\%$ \\
            & $6$ & $\mathrm{W}_2/\mathrm{S}_2$ & Sixth nearest neighbor, $\SI{5.5}{\AA} \pm 5\%$ \\
    \addlinespace
    $m = 3$ & $(1,1,2)$ & $\mathrm{WS}_2/\mathrm{W}_2\mathrm{S}$ & 2x 1nn + 1x 2nn isosceles triangle \\
            & $(2,2,2)$ & $\mathrm{W}_3/\mathrm{S}_3$ & 3x 2nn equilateral triangle \\
            & $(1,2,3)$ & $\mathrm{WS}_2/\mathrm{W}_2\mathrm{S}$ & 1nn + 2nn + 3nn scalene triangle \\
            & $(2,3,3)$ & $\mathrm{WS}_2$ & 1x 2nn + 2x 3nn isosceles triangle \\
    \addlinespace
    $m = 4$ & $(1,1,1,1,2,2)$ & $\mathrm{W}_2\mathrm{S}_2$ & 4x 1nn + 2x 2nn tetrahedron \\
            & $(1,1,1,2,2,2)$ & $\mathrm{W}_3\mathrm{S}/\mathrm{WS}_3$ & 3x 1nn + 3x 2nn tetrahedron \\
            & $(1,1,2,2,2,3)$ & $\mathrm{W}_3\mathrm{S}/\mathrm{WS}_3$ & 2x 1nn + 3x 2nn + 1x 3nn tetrahedron \\
    \bottomrule
    \end{tabularx}
    \captionof{table}{Description of the cluster set used to compute the feature vector in Eq.~\eqref{eq:tce-feat-vec}, visualized in Fig.~\ref{fig:ws2-clusters}.}
    \label{tab:mx2-many-body-clusters}

\end{table*}

We note that each $(m,\ell)$ pair has a finite set of permissible species within the cluster. This is due to the fact that $\text{W}\text{S}_2$ spans multiple sublattices. As an example, the aforementioned $(m, \ell) = (2, 1)$ cluster can only describe interactions between the metal and chalcogen sublattices.

Within \verb|tce-lib|, the only inputs to define such a feature set are the labels $\ell$ and the nearest neighbor-distances. Many features are identically $0$, however, as described above. This is not problematic for using feature vectors as global descriptors, as done in this work, but can be problematic when fitting a TCE model on systems with multiple sublattices. In the example described above, $N_{\text{W}\text{W}}^{(1)}$ is identically $0$, so that any value of $\varepsilon_{\text{W}\text{W}}^{(1)}$ is valid.

\subsection{TCE fitting and MC runs}\label{sec:tce+mc}

Using the feature set $\Lambda$ described in Fig.~\ref{fig:ws2-clusters}, we fit a TCE model using $L^1$ (Lasso) regression using \verb|scikit-learn| \cite{scikit-learn, sklearn_api}, minimizing the loss:

\begin{equation}
    \mathcal{L}(\boldsymbol{\varepsilon}) = \frac{1}{2N}\sum_{s=1}^N \left ( E_s - \boldsymbol{\varepsilon}:\mathbf{N}_s\right )^2 + \rho \left\| \boldsymbol{\varepsilon} \right\|_{\mathbf{1}, \boldsymbol{\varphi}}
\end{equation}

where $s$ indexes samples, $N$ is the number of training samples, $E_s$ is the relaxed DFT energy of the $s$'th sample, $\boldsymbol{\varepsilon}$ are the interaction coefficients (which are learned parameters), $\mathbf{N}_s$ is the number of $[\ell]$- and $\boldsymbol{\alpha}$-type clusters in sample $s$, $:$ denotes the tensor dot product, $\rho$ is a regularization parameter, and $\|\cdot\|_{\mathbf{1}, \boldsymbol{\varphi}}$ denotes a generalized $L^1$ tensor norm equipped with a linear transformation $\boldsymbol{\varphi}$:

\begin{equation}
    \begin{aligned}
        X_{si\alpha} &= \begin{cases}
            1 & \text{site $i$ in sample $s$ has type $\alpha$} \\
            0 & \text{else}
        \end{cases} \\
        N_{s\alpha_1\cdots\alpha_m}^{[\ell]} &= A_{i_1\cdots i_m}^{[\ell]}\prod_{k=1}^m X_{si_k\alpha_k} \\
        \boldsymbol{\varepsilon}:\mathbf{N}_s &= \sum_{[\ell]\in\Lambda} \sum_{\alpha_1\cdots\alpha_m} \varepsilon_{\alpha_1\cdots\alpha_m}^{[\ell]} N_{s\alpha_1\cdots\alpha_m}^{[\ell]} \\
        \left\| \boldsymbol{\varepsilon}\right\|_{\mathbf{1}, \boldsymbol{\varphi}} &= \sum_{[\ell]\in\Lambda}\sum_{\alpha_1\cdots\alpha_m}\left|\varphi_{\alpha_1\cdots\alpha_m, \alpha_1' \cdots\alpha_m'}^{[\ell], [\ell']}\varepsilon_{\alpha_1'\cdots\alpha_m'}^{[\ell']}\right|
    \end{aligned}
\end{equation}

The loss above is minimized over a training set. Within this training set, we perform $5$-fold cross validation and pick the $\rho$ that minimizes the loss. Furthermore, to promote stability, training features are rescaled to have zero mean and unit standard deviation before fitting. This rescaling defines the linear transformation $\boldsymbol{\varphi}$ as the inverse of the rescaling transformation $\boldsymbol{N}_s\mapsto \Tilde{\boldsymbol{N}}_s$, so that the tensor product $\boldsymbol{\varepsilon} : \mathbf{N}_s = (\boldsymbol{\varphi}\boldsymbol{\varepsilon}):\Tilde{\mathbf{N}}_s$ is invariant under the transformation. Here, we choose $L^1$ regularization to promote sparsity in the $\boldsymbol{\varepsilon}$ coefficients. The best-fit $\boldsymbol{\varepsilon}$ is indeed very sparse, i.e. only about $3\%$ of coefficients are nonzero.

\begin{figure}[H]
    \centering
    \includegraphics[width=\linewidth]{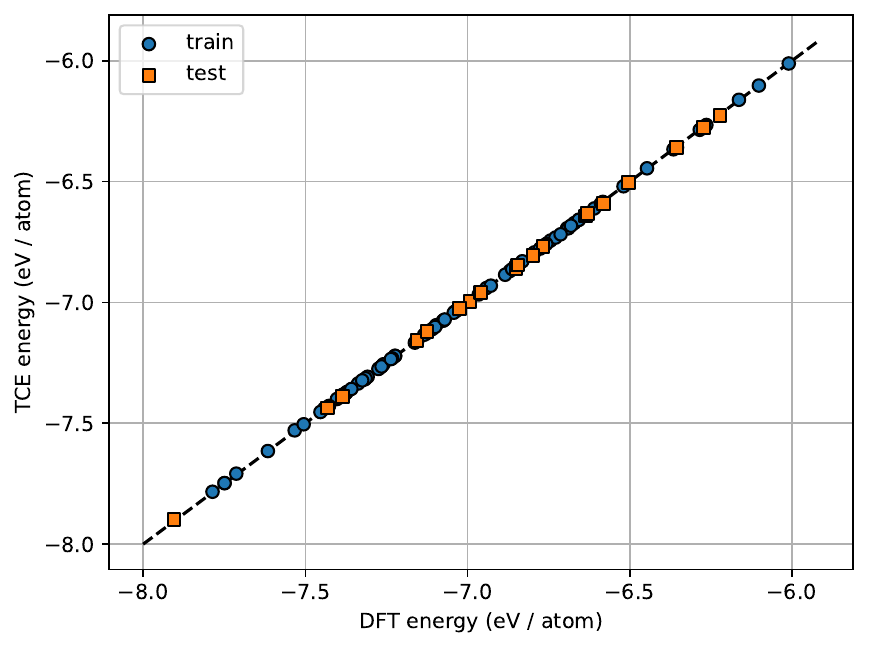}
    \caption{Parity plot for the fitted TCE model with respect to the relaxed DFT energies.}
    \label{fig:parity2}
\end{figure}

We find that the fit is excellent at predicting the raw DFT energies within both the training and test set (Fig.~\ref{fig:parity2}).

Then, using the fitted model, we run canonical Monte Carlo on the $M(X, X')_2$ monolayers for $100,000$ steps at varying temperatures with system sizes of $15\times 15$, and characterize the chemical ordering via the Warren-Cowley short range order parameters $\chi_{\alpha\beta}$ \cite{cowley1950approximate}:

\begin{equation}
    \chi_{\alpha\beta} = 1 - \frac{p_{\alpha\beta}}{(2-\delta_{\alpha\beta})x_\alpha x_\beta}
\end{equation}

where $\alpha$ and $\beta$ denote alloying types, $p_{\alpha\beta}$ is the probability that a bond within the shell is of type $\alpha$-$\beta$, $x_\alpha$ is the concentration of type $\alpha$, and $\delta_{\alpha\beta}$ is the Kronecker delta. Here, $\chi_{\alpha\beta} = 0$ denotes full chemical disorder within the shell, $\chi_{\alpha\beta} < 0$ denotes $\alpha$-$\beta$ clustering, and $\chi_{\alpha\beta} > 0$ denotes $\alpha$-$\beta$ repulsion. We compute the average SRO parameter only within the chalcogen sublattice, i.e. $M$-$X$ interactions are not counted.

\begin{figure}[H]
    \centering
    \includegraphics[width=\linewidth]{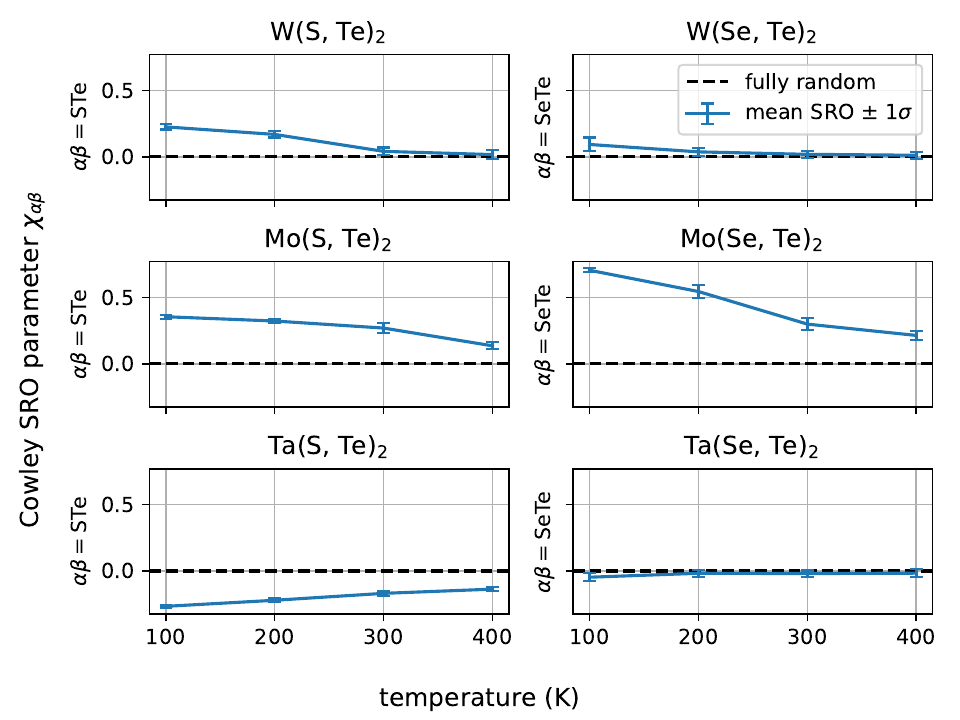}
    \caption{Warren-Cowley short range order parameters equilibrated via Monte Carlo as a function of temperature $T$.}
    \label{fig:wc-sros}
\end{figure}

As expected, we find that the SRO parameters tend to zero at high temperatures. We find that, for the $\text{W}(X, X')_2$ and $\text{Mo}(X, X')_2$ systems, that $X$-$X$ clustering are favorable at low temperatures. Visualizations for each TMD at the two extreme temperatures are included in Fig.~\ref{fig:visualization}, where the clustering behavior is more obvious. Interestingly, this behavior is not preserved for either $\text{Ta}(X, X')_2$ system. Namely, $\text{Ta}(\text{S}, \text{Te})_2$ favors $X$-$X'$ clustering at low temperatures, and $\text{Ta}(\text{Se},\text{Te})_2$ remains essentially random for the entire tested temperature range. 

\subsection{Raw VACF plots}\label{sec:vacf}

Below we show the raw VACF data $C(t)$ defined in Eq.~\eqref{eq:vacf-and-vdos}.

\begin{figure}[H]
    \centering
    \includegraphics[width=\linewidth]{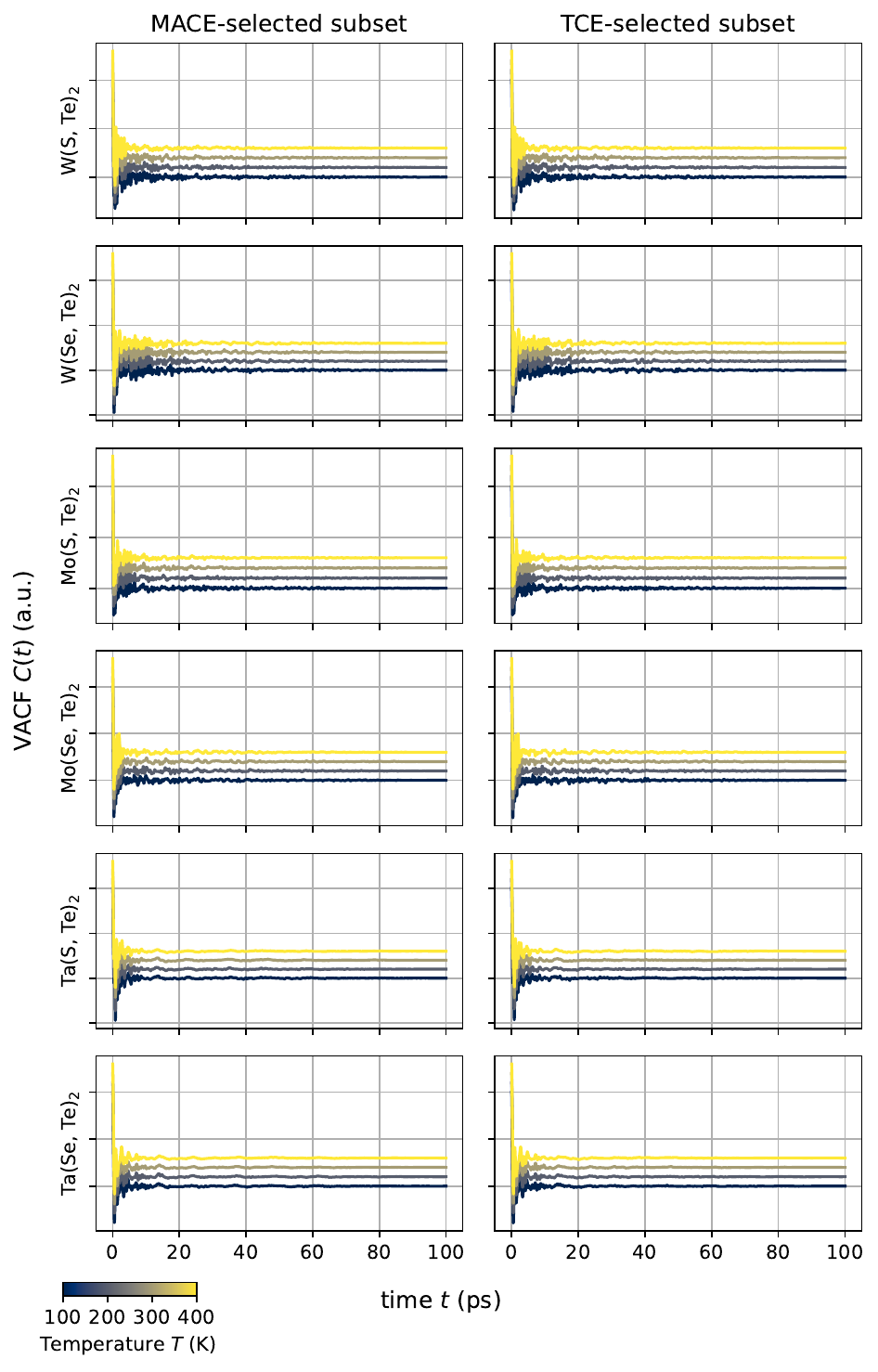}
    \caption{VACF $C(t)$ for all $M(X,X')_2$ monolayers as a function of temperature for the MACE-selected- (left) and TCE-selected- (right) subset finetuned MLIPs.}
    \label{fig:vacf}
\end{figure}

The important feature of the above plot is that, in all systems, the correlations vanish over the MD simulation time, meaning that each system is reasonably equilibrated with respect to atomic vibrations.